\documentclass[pra,twocolumn,superscriptaddress,showpacs]{revtex4}

\usepackage[dvips]{graphics}
\usepackage{amssymb}
\usepackage{amsmath}
\usepackage{epsfig}
\usepackage{latexsym}
\usepackage{color}
\usepackage{rotating}
\usepackage{float}

\newcommand{\cosec}{\mathrm{cosec} \,}

\begin{document}

\title{Two-mode back-action-evading measurements in cavity optomechanics}

\author{M. J. Woolley} 
\affiliation{Department of Physics, McGill University, Montr\'{e}al, QC, H3A 2T8, Canada}
\affiliation{School of Engineering and Information Technology, UNSW Canberra, Canberra, ACT, 2600, Australia}
\author{A. A. Clerk}
\affiliation{Department of Physics, McGill University, Montr\'{e}al, QC, H3A 2T8, Canada}

\begin{abstract}
We study theoretically a three-mode optomechanical system where two mechanical oscillators are coupled to a single cavity mode.  By using two-tone (i.e.~amplitude-modulated) driving of the cavity, it is possible to couple the cavity to a single collective quadrature of the mechanical oscillators. In such a way, a back-action-evading measurement of the collective mechanical quadrature is possible.  We discuss how this can allow one to measure both quadratures of a mechanical force beyond the full quantum limit, paying close attention to the role of dissipation and experimental imperfections.  We also describe how this scheme allows one to generate steady-state mechanical entanglement; namely, one can conditionally prepare an entangled, two-mode squeezed mechanical state.  This entanglement can be verified directly from the measurement record by applying a generalized version of Duan's inequality; we also discuss how feedback can be used to produce unconditional entanglement.
\end{abstract}


\maketitle


\section{Introduction}
\label{sec:intro}

The idea of a back-action-evading quantum measurement is by now a well-known concept \cite{Braginsky80,caves:RMP,TMSS,BraginskyBook}. The simplest and most studied realization is the continuous position measurement of a simple harmonic oscillator. If one tries to simultaneously follow both quadratures of the oscillator's motion (i.e.~both the amplitudes of the  sine and cosine components of the motion), then the unavoidable effects of quantum back-action imply that one cannot improve the measurement indefinitely by increasing the measurement strength. This leads to the quantum limit on  continuous position detection \cite{BraginskyBook,clerk:review}. If instead one measures just a single quadrature of the motion (say the $\hat{X}$ quadrature), there is no such limit as one has ``evaded" the back-action:  the measurement back-action only heats the conjugate $\hat{P}$ quadrature, which has no effect on the measurement as the dynamics does not couple it to $\hat{X}$.  

Such back-action-evading (BAE) measurement schemes allow one to measure one quadrature of a narrow-bandwidth force acting on a mechanical oscillator with arbitrary precision.  They also naturally lead to squeezing of the measured quadrature, albeit a conditional squeezing, where one must have access to the full measurement record to produce unconditional squeezing. Back-action-evading schemes are also known  as  ``quantum-non-demolition (QND) in time" measurements, as one is making a QND measurement of an observable which is explicitly time-dependent.  These ideas have recently been discussed \cite{clerk:QND} 
within the specific context of quantum optomechanics \cite{woolleyreview}, where a mechanical oscillator is dispersively coupled to a driven cavity.  A back-action-evading single-quadrature measurement was even implemented experimentally by Herzberg \emph{et al.}, using a nanomechanical oscillator coupled to a driven microwave-frequency cavity \cite{schwab}.

One could now ask about exploiting the idea of back-action-evasion in more general settings.  The simplest generalization would involve two harmonic oscillators, implying four quadratures of motion.  Such a two-mode BAE scheme would involve measuring two quadratures of motion, with the back-action only driving the unmeasured conjugate quadratures. If one could suitably couple the system, one could imagine measuring {\it both} quadratures of an applied force without any quantum limit.

\setlength{\abovecaptionskip}{-10pt }

\begin{figure}[H]
\begin{center}
\includegraphics[scale=0.43]{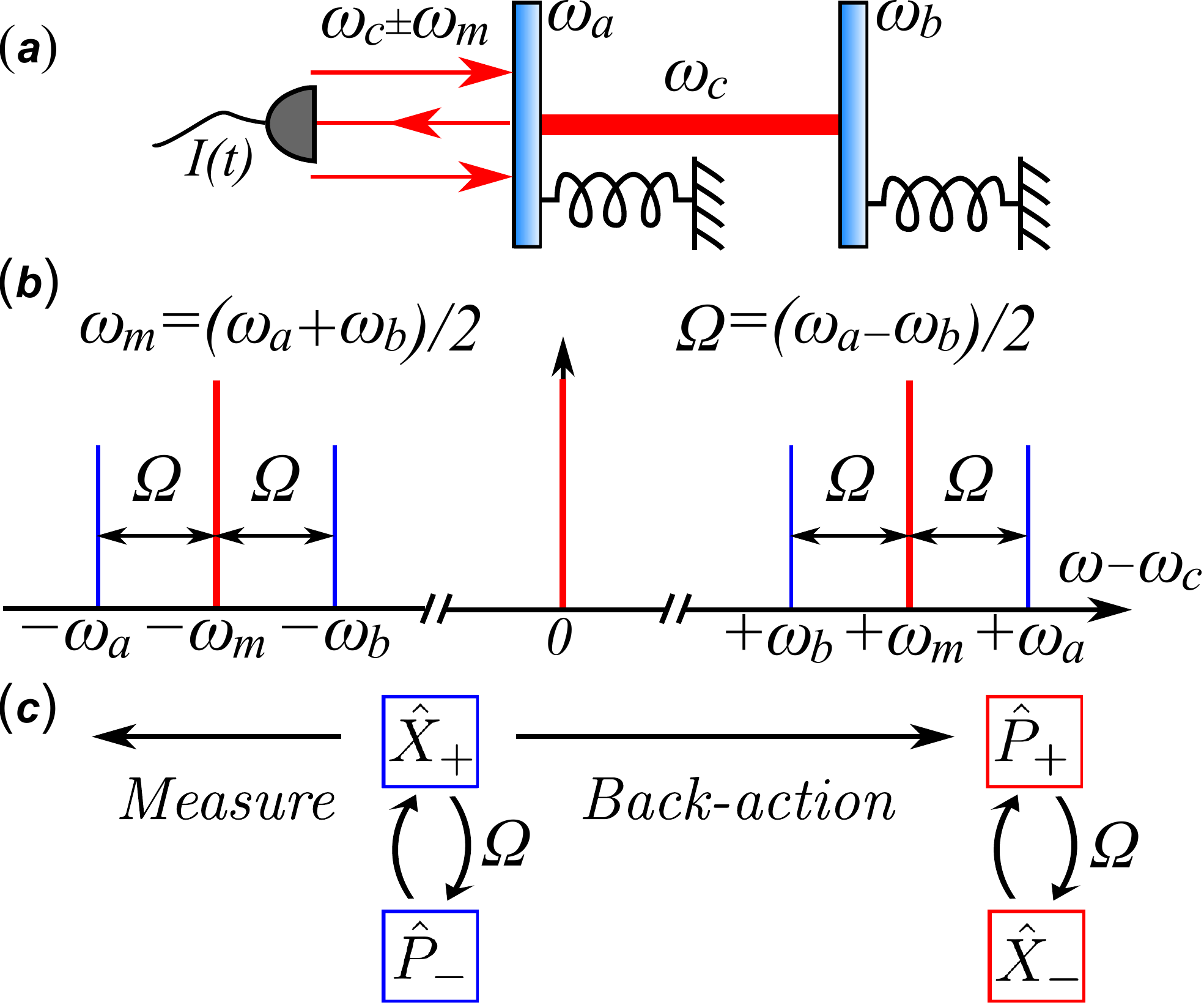}
\end{center}
\caption{ (Color online). (a) The system studied consists of two mechanical oscillators, each coupled to a cavity (or circuit) mode. Measurement is performed via phase-sensitive detection near the cavity resonance of the cavity output field. (b) Frequencies in this system, defined with respect to the cavity resonance frequency $\omega_c$. The blue lines indicate the standard mechanical sidebands, at $\pm \omega_a$ and $\pm \omega_b$. 
The cavity is driven at $\pm \omega_m = \pm ( \omega_a + \omega_b )/2$, as indicated by vertical red lines on either side of the cavity resonance. 
(c) Representation of the two-mode BAE scheme in the ideal, fully symmetric case. The observable $\hat{X}_+$ is directly measured, and the observable $\hat{P}_+$ is heated by the corresponding back-action. $\hat{P}_-$ is dynamically coupled to $\hat{X}_+$, and therefore effectively measured provided that $\Omega \gg \gamma$. In this limit, $\hat{X}_-$ (which is dynamically coupled to $\hat{P}_-$) is also heated by the back-action of the measurement. } 
\label{fig:optodualmechanics}
\end{figure}


That such a scheme is possible was recently discussed in detail by Tsang and Caves \cite{tsang:QNC,tsang:evadeQM}, though similar ideas have been discussed earlier (see, for example, Ref.~\onlinecite{koopman} or App.~D of Ref.~\onlinecite{james}).  From a practical perspective, it appears challenging, as it requires one of the two oscillators involved to have effectively a negative mass (and hence a negative frequency).  Similar to standard single-oscillator back-action-evasion, one might expect that there is also a squeezing aspect to such two-mode BAE.  This was discussed by Hammerer \emph{et al.}\,\cite{hammerer:PRL}:  the squeezing here of the delocalized measured quadratures naturally gives rise to an EPR-style entanglement between the two modes. 

In this paper, we now extend the above idea of two-mode back-action-evasion to quantum optomechanics, specifically considering a system where two mechanical oscillators are coupled to a single electromagnetic mode, as depicted in Fig.~\ref{fig:optodualmechanics}.  Such two-mode optomechanical systems have recently been realized experimentally \cite{sillanpaa}.  We show that the two-mode BAE scheme can be realized in a relatively simple fashion in such a system, using a generalization of the two-tone driving scheme used to achieve single-mode back-action-evasion \cite{Braginsky80,clerk:QND}.  This is in contrast to previous proposals, where to achieve the requisite negative mass, one of the two modes was not mechanical, but rather an atomic ensemble \cite{hammerer:PRL} or another cavity mode \cite{tsang:QNC}. Unlike the single-mode back-action-evasion case, we expect that this scheme will \emph{not} be susceptible to parametric instabilities \cite{suh:instabilities}, since here the number of photons in the cavity will \emph{not} oscillate at double the mechanical resonance frequencies. 

Note that applications of BAE measurement ideas to a system of two mechanical oscillators have previously been studied \cite{bocko:QND,bocko:QND1,onofrio:QNDmodulated,onofrio:QNDstroboscopic,rioli:stroboscopicsqueezing,cinquegrana:stroboscopicSNR,bonifazi}, motivated by attempts at gravitational wave detection. The approaches presented in those works differ significantly from our scheme; in particular, they involve an explicit coupling between the mechanical oscillators, something that is not required with our scheme.


We provide a thorough discussion of the optomechanical two-mode BAE scheme, including a discussion of expected experimental imperfections, and methods for countering these.  In addition, we provide a full analysis of the force sensitivity of this scheme and the possibility of beating conventional quantum limits. Unlike previous work \cite{tsang:QNC,tsang:evadeQM}, we include dissipation in this analysis. 

Finally, we also consider the generation of stationary, conditional entanglement using this scheme, and the possibility of turning this into unconditional entanglement using feedback.  This is in contrast to more conventional measurement-based entanglement generation schemes, which rely on strong measurements and post-selection; such schemes are well-established in optics \cite{zeilinger,kimble,kuzmich}, and have recently been discussed in the context of superconducting circuits \cite{blais,johansson}.  Our scheme also differs from that in Ref.\,\onlinecite{hammerer:PRL} as we generate \emph{stationary} entanglement, and do not require a single-shot strong feedback operation. 

The remainder of this paper is structured as follows. In Sec.~\ref{sec:system} we derive the Hamiltonian and Heisenberg-Langevin equations describing this system and introduce a method for compensating for system asymmetries. In Sec.~\ref{back-action} we calculate the back-action heating effect and the measured noise spectrum, with a view to assessing the force sensing limits of our system. 

In Sec.~\ref{sec:forcesensitivity} we rigorously compare the force sensitivity of our scheme, including imperfections, against conventional quantum limits. The situation is significantly more complicated when one includes the presence of dissipation (unlike Refs.\,\onlinecite{tsang:QNC} and \onlinecite{tsang:evadeQM}). We find that for a signal force applied at the mechanical resonance frequency, the number of added noise quanta can be made to go to zero (below the quantum limit of $1/2$),
\begin{equation}
\bar{n}_{\rm add} [\omega_a ] \rightarrow 0 ,
\end{equation}
even in the presence of dissipation. However, this is only possible in a narrow bandwidth about the mechanical resonance and requires a careful tuning of optomechanical couplings. For a signal force off-resonant from the mechanical resonance frequency (detuned by an amount $\Delta$), we find that the added noise can be made to go to
\begin{equation}
\bar{n}_{\rm add} [\omega_a + \Delta ] = \frac{1}{2\sqrt{2}} ,
\end{equation}
well below the standard quantum limit in this case ($\bar{n}_{\rm add}[\omega_a + \Delta ] = \Delta/\gamma$ where $\gamma$ is the average mechanical damping rate) and even below the full quantum limit (i.e. allowing for optimal detector noise correlations).

The conditional dynamics of the mechanical oscillators, under a continuous measurement via the coupled cavity, are described in Sec.~\ref{sec:conditionalvariances}. The steady-state conditional variances are determined, and these are used to determine in which regimes the mechanical oscillators are conditionally entangled. The presence of all-mechanical entanglement is determined from Duan's inequality \cite{duan:inseparability}, which here (in the case of symmetric optomechanical couplings and mechanical damping rates) becomes the requirement
\begin{equation}
V_{X_+} + V_{P_-} = \sqrt{ \frac{2( \bar{n}_{\rm th} + 1/2 )}{\eta C} } < 1 , \label{eq:DuanIntro}
\end{equation}
where $V_{X_+} \ (V_{P_-})$ is the variance in the sum (difference) of the mechanical $\hat{X} \ (\hat{P})$ quadratures, $\bar{n}_{\rm th}$ is the effective thermal occupation of the mechanical environment (see Eq.~(\ref{eq:occupations1})), $\eta$ is the quantum efficiency of the measurement (see Eq.~(\ref{eq:fullME})), and $C$ is the so-called optomechanical cooperativity, essentially quantifying the measurement strength (see Eq.~(\ref{eq:cooperativity})). Both quantities on the left-hand-side of Eq.~(\ref{eq:DuanIntro}) may be obtained in a straightforward manner from the measurement record. Therefore, the mechanical oscillators should be entangled provided that the cooperativity obeys
\begin{equation}
C > \frac{ 2( \bar{n}_{\rm th} + 1/2) }{\eta} .
\end{equation}
From an experimental perspective, this would appear to be a modest requirement, well within the reach of current optomechanics experiments based on microwave circuits \cite{teufel:groundstate}. In Sec.~\ref{sec:unconditionalvariances}, the total unconditional variances of the collective quadratures are calculated, and a feedback scheme is introduced that allows the conditional entanglement to be converted to unconditional entanglement.  
   



\section{System}
\label{sec:system}

\subsection{Hamiltonian}
The system (see Fig.~\ref{fig:optodualmechanics}) is composed of two mechanical oscillators, with resonance frequencies $\omega_a$ and $\omega_b$, each independently and dispersively coupled (with strengths $g_a$ and $g_b$, respectively) to a common cavity mode having resonance frequency $\omega_c$. The Hamiltonian is
\begin{eqnarray}
\mathcal{\hat{H}} & = & \omega_a \hat{a}^\dagger \hat{a} + \omega_b \hat{b}^\dagger \hat{b} + \omega_c \hat{c}^\dagger \hat{c} + g_a \left( \hat{a}+\hat{a}^\dagger \right) \hat{c}^\dagger \hat{c}  \nonumber \\ 
& & + g_b ( \hat{b}+\hat{b}^\dagger ) \hat{c}^\dagger \hat{c} + \hat{H}_{\rm drive} + \hat{H}_{\rm diss} , \label{eq:HamOne}
\end{eqnarray}
where $\hat{a}$ and $\hat{b}$ denote mechanical mode lowering operators, $\hat{c}$ denotes the electromagnetic mode lowering operator, and $\hat{H}_{\rm drive}$ accounts for driving of the electromagnetic mode. The term $\hat{H}_{\rm diss}$ accounts for dissipation, with the modes subject to damping at rates $\gamma_a$, $\gamma_b$ and $\kappa$, respectively. 

To realize the two-mode back-action-evading dynamics suggested by Tsang and Caves \cite{tsang:QNC}, we need the two mechanical oscillators to have equal and opposite frequencies, and moreover, have the cavity (the detector in this scheme) only couple to a single collective quadrature of the two mechanical oscillators (an operator which is explicitly time-dependent).  Both these requirements can be realized by simply adapting the two-tone (i.e. ampltiude-modulated) driving scheme used for ordinary single-quadrature BAE in Refs.~\onlinecite{clerk:QND} and \onlinecite{schwab}.  One applies a drive to the cavity at both $\omega_c + \omega_m$ and $\omega_c - \omega_m$, where $\omega_m = (\omega_a + \omega_b)/2$ is the average of the two mechanical frequencies, i.e.:
\begin{eqnarray}
	\hat{H}_{\rm drive} & = & \left( \mathcal{E}^*_+ e^{-i \omega_m t} + \mathcal{E}^*_- e^{i \omega_m t} \right) e^{i \omega_c t} \hat{c} + \textrm{h.c.} \label{eq:driving}
\end{eqnarray}  
Working in an interaction picture with respect to $\hat{H}_0 =  \omega_m \left( \hat{a}^\dagger \hat{a} +  \hat{b}^\dagger \hat{b} \right) + \omega_c \hat{c}^\dagger \hat{c}$, one finds that the frequencies of the mechanical oscillators are now $\pm \Omega$ as desired, where $2\Omega = \omega_a - \omega_b$.  We further specialize to 
an optomechanical system in the good cavity limit ($\omega_m \gg \kappa$) and assume large driving amplitudes $| \mathcal{E}_+ |,|\mathcal{E}_-|$, such that one can linearize the optomechanical interaction.
In this standard regime, one finds (as desired) that in the interaction picture, the cavity only couples to a collective mechanical quadrature with a time-independent coupling.  Specifically, defining the the quadrature operators for each mode in the interaction picture in the standard manner ($\alpha=a,b,c$):
\begin{subequations}
\begin{eqnarray}
	\hat{X}_\alpha & = & (\hat{\alpha}+\hat{\alpha}^\dagger)/\sqrt{2}, \\
	\hat{P}_\alpha & = & -i (\hat{\alpha} - \hat{\alpha}^\dagger)/\sqrt{2},
\end{eqnarray}
\end{subequations}
and collective (canonically conjugate) mechanical quadrature observables via
\begin{subequations}
\begin{eqnarray}
	\hat{X}_\pm & \equiv & (\hat{X}_a \pm \hat{X}_b )/ \sqrt{2}, \\
	\hat{P}_\pm & \equiv & (\hat{P}_a \pm \hat{P}_b)/ \sqrt{2} ,
\end{eqnarray}
\end{subequations}
the optomechanical Hamiltonian takes the form
\begin{eqnarray}
\mathcal{\hat{H}} & =  & \Omega \left( \hat{X}_+ \hat{X}_- + \hat{P}_+ \hat{P}_- \right) + G\hat{X}_+ \hat{X}_c - G_d \hat{X}_- \hat{X}_c \nonumber \\ 
& & + \hat{H}_{\rm diss}. \label{eq:HamUntilded} 
\end{eqnarray}
Here, the many-photon optomechanical couplings are given by
\begin{subequations}
\begin{eqnarray}
G & \equiv & \sqrt{2} \left( g_a + g_b \right) \bar{c} , \\ 
G_d & \equiv & \sqrt{2} \left( g_b - g_a \right)  \bar{c} , \label{eq:asymcoupling}
\end{eqnarray}
\end{subequations}
where $\bar{c} =  |\mathcal{E}_\pm | /  \sqrt{ \omega^2_m + \kappa^2/4 } $ 
is the steady-state amplitude of the cavity field at the driven sidebands. Note that we want to be in a regime where $\kappa \gtrsim \Omega$, such that the cavity responds fast enough to monitor the collective mechanical oscillation. The details of this derivation are given in App.~\ref{HamiltonianDerivation}.  We stress here that the interaction terms in Eq.~(\ref{eq:HamUntilded}), even accounting for contributions from the \emph{off-resonant} sidebands of the \emph{driving} fields, will \emph{not} include terms oscillating at double the mechanical resonance frequencies. Accordingly, the cavity photon number will not oscillate at this frequency and the parametric instabilities associated with single-mode back-action-evasion \cite{suh:instabilities} should be avoided in the two-mode case. 

For equal optomechanical couplings (i.e.~$G_d = 0$) and equal mechanical damping rates 
($\gamma_a = \gamma_b \equiv \gamma$) this system
perfectly realizes the BAE scheme of Tsang and Caves.  To see this, consider the Heisenberg-Langevin equations
for our system in this ideal symmetric limit.  Defining the vector of collective mechanical quadrature operators as
\begin{equation}
\vec{V} = (\hat{X}_+, \hat{P}_-, \hat{X}_-, \hat{P}_+)^T, \label{eq:vectoroperators}
\end{equation}
a standard calculation yields
\begin{eqnarray}
	\frac{d}{dt} \vec{V} = \mathbf{M} \cdot \vec{V} + \vec{F}_{\rm BA} + \mathbf{N} \cdot \vec{\xi}. 
	\label{eq:QLE}
\end{eqnarray}
Here, $\mathbf{M}$ describes the oscillator dynamics in the rotating frame, 
\begin{eqnarray}
	\mathbf{M} = 
\left[
\begin{array}{cccc}
	- \gamma/2	&	\Omega	&	0	&	0   \\
  	-\Omega			&	 -\gamma/2  &	0	&	0   \\
	0	&	0	&	-\gamma/2	&	\Omega	   \\
  	0	&	0	&	-\Omega			&	 -\gamma/2     \\
\end{array}
\right], \label{eq:MDefs}
\end{eqnarray}
while $\vec{F}_{\rm BA}$ is the back-action force from the cavity:
\begin{equation}
	\vec{F}_{\rm BA} = 
		( 0, 0,0,-G \hat{X}_c )^T . \label{eq:FDefs}
\end{equation}
Further,
\begin{equation}
\vec{\xi} = (\hat{X}_{+,{\rm in}}, \hat{P}_{-,{\rm in}}, \hat{X}_{-,{\rm in}}, \hat{P}_{+,{\rm in}})^T, \ \ \ \mathbf{N} = \sqrt{\gamma} \mathbf{I}_4 , \label{eq:noiseinput}
\end{equation}
describe the Langevin noise arising from the dissipative baths of each mechanical resonator, with $\mathbf{I}_n$ denoting the $n \times n$ identity matrix.  Finally, the equations of motion for the cavity quadratures are:
\begin{subequations}
\begin{eqnarray}
	\dot{\hat{X}}_c & = & -\frac{\kappa}{2} \hat{X}_c + \sqrt{ \kappa } \hat{X}_{c,{\rm in}} , \ \label{eq:Xcav}  \\
	\dot{\hat{P}}_c & = & - G \hat{X}_+ - \frac{\kappa}{2} \hat{P}_c + \sqrt{ \kappa } \hat{P}_{c,{\rm in}} . \label{eq:QLEPcav}   
\end{eqnarray}
\end{subequations}
We thus have the desired behaviour, namely:
\begin{itemize}
\item
	The collective $\hat{X}_+$ and $\hat{P}_-$ quadratures are linked dynamically the same way as $\hat{X}$ and $\hat{P}$ for a harmonic oscillator, but are dynamically
	independent from their conjugate pair ($\hat{P}_+$ and $\hat{X}_-$).  In the language of Tsang and Caves, 
	the commuting observables $\hat{X}_+$ and $\hat{P}_-$ form a quantum-mechanics-free-subspace.
\item
	From Eq.~(\ref{eq:QLEPcav}), we see that the cavity measures $\hat{X}_+$ (and hence the dynamically linked $\hat{P}_-$ quadrature), and its back-action only drives the unmeasured mechanical quadratures $\hat{P}_+$ and $\hat{X}_-$ (c.f.~Eq.~(\ref{eq:FDefs})).
\end{itemize}
One thus has a setup which allows the back-action-evading measurement of both the collective quadratures $\hat{X}_+$ and $\hat{P}_-$. Information about the motion of these collective quadratures appears at the cavity output at sidebands $\pm \Omega$ from the cavity resonance.

\subsection{Compensating for unequal optomechanical couplings}
\label{subsec:Asymmetry}

While the scheme is easy to understand in the perfectly symmetric case, any real expeirment will have to contend with both an asymmetry in the optomechanical couplings ($g_a \neq g_b$) and an asymmetry in the mechanical damping ($\gamma_a \neq \gamma_b$).  Asymmetric couplings cause $G_d$ in Eq.~(\ref{eq:HamUntilded}) to be non-zero, implying that the
cavity measures both the $\hat{X}_+$ and $\hat{X}_-$ collective quadratures.  As a result, the measurement is no longer back-action-evadig (i.e.~the unwanted measurement of $\hat{X}_-$ causes back-action to drive $\hat{P}_-$, which then corrupts the measurement of $\hat{X}_+$).  Even if there is no coupling asymmetry, if $\gamma_a \neq \gamma_b$, the damping terms in the Heisenberg-Langevin equations will link 
$\hat{X}_+$ and $\hat{X}_-$, also ruining the back-action-evasion.  Both these effects can easily be seen by writing out the Heisenberg-Langevin equations in the general asymmetric case; see App.~\ref{sec:HLEquations}, especially Eqs.~(\ref{eq:Moriginalbasis}) and (\ref{eq:BAoriginalbasis}).  

We will show in later sections that while small deviations from the ideal symmetric system cause a departure from perfect back-action-evasion, one can still beat quantum limits on detection as well as generate squeezing and entanglement.  Here, we point out that it is possible to {\it exactly} null the deleterious effects of asymmetric optomechanical couplings
by introducing additional parametric drives on each mechanical resonator. Note that attempting to compensate for the asymmetric optomechanical couplings by modifying the weights of the two cavity drives only leads to more unwanted interaction terms. 

To begin our analysis of the case $g_a \neq g_b$, it is useful to introduce a new set of canonically-conjugate collective mechanical quadratures, such that in this basis, the cavity only couples to a single (collective) quadrature.  Defining the rotation angle $\theta$ as:
\begin{equation}
	\theta = \tan^{-1}\left( G_d/G \right),
	\label{eq:xiDefn}
\end{equation}
we introduce:
\begin{subequations}
\begin{eqnarray}
\tilde{X}_\pm & \equiv & \cos \theta \ \hat{X}_\pm \mp \sin \theta \ \hat{X}_\mp, \label{eq:tilded0} \\ 
\tilde{P}_\pm & \equiv  & \cos \theta \ \hat{P}_\pm \mp \sin \theta \ \hat{P}_\mp , \label{eq:tilded}.
\end{eqnarray}
\end{subequations}
The Hamiltonian of Eq.~(\ref{eq:HamUntilded}), in terms of the observables introduced in Eqs.~(\ref{eq:tilded0}) and (\ref{eq:tilded}), takes the form
\begin{eqnarray}
	\hat{\mathcal{ H }} & = & \tilde{\Omega} \left( \tilde{X}_+ \tilde{X}_- + \tilde{P}_+ \tilde{P}_- \right) + \tilde{G} \tilde{X}_+ \hat{X}_c  + \hat{H}_{\rm diss} \nonumber \\
	& & + \frac{1}{2}p\tilde{\Omega} \left( \tilde{X}^2_- + \tilde{P}^2_- - \tilde{X}^2_+ - \tilde{P}^2_+ \right), 
	\label{eq:Hamtilded}
\end{eqnarray}
where we have defined:
\begin{subequations}
\begin{eqnarray}
p & \equiv & \tan 2\theta = \frac{2G_d/G}{ 1-G^2_d/G^2 } = \frac{g_a^2 - g_b^2}{2 g_a g_b} , \label{eq:pDefn} \\
\tilde{\Omega} & \equiv & \Omega \cos 2\theta = \Omega \frac{ 1 - G^2_d/G^2 }{ 1+G^2_d/G^2 } = 2 \frac{g_a  g_b^2}{g_a^2 +  g_b^2}, \label{eq:tildeOmegaDefn} \\ 
\tilde{G} & \equiv & G \sec \theta = G \sqrt{1+G^2_d/G^2} . 
	\label{eq:GTilde}
\end{eqnarray}
\end{subequations}
One can recast the Heisenberg-Langevin equations describing the system in terms of these rotated observables. The system retains the form of Eq.~(\ref{eq:QLE}), now in terms of the vector of rotated observables,
\begin{equation}
\vec{V} = ( \tilde{X}_+, \tilde{P}_-, \tilde{X}_-, \tilde{P}_+ )^T , \label{eq:rotatedvector}
\end{equation} 
with the input noise vector given by
\begin{equation}
\vec{\xi} = ( \tilde{X}_{+,{\rm in}} , \tilde{P}_{-,{\rm in}} , \tilde{X}_{-,{\rm in}} , \tilde{P}_{+,{\rm in}} )^T . \label{eq:tildenoise}
\end{equation}
The appropriate matrices are specified in Eqs.~(\ref{eq:rotatedsystem}) and (\ref{eq:tildeBA}).

The cavity is now only coupled to a single collective quadrature $\tilde{X}_+$, with a modified coupling constant $\tilde{G}$.  
The free evolution of the oscillators in this quadrature basis is, however, not of the desired form; the back-action-evasion is ruined by the terms in the last line of Eq.~(\ref{eq:Hamtilded}), which 
dynamically link the measured  observables $(\tilde{X}_+,\tilde{P}_-)$ to their (perturbed) conjugate pair  $(\tilde{P}_+,\tilde{X}_-)$

It is now natural to ask whether it is possible to modify the mechanical system in some simple way so as to eliminate the last line in Eq.~(\ref{eq:Hamtilded}).  Such a modification would require introducing both position and momentum couplings between the two mechanical oscillators, something that is not experimentally feasible.  A more modest approach would involve simply adding terms to the Hamiltonian of each mechanical oscillator (but not introduce any coupling).  We consider parametrically driving each mechanical oscillator, and thus adding a term $\hat{H}_{\rm pa}$ to the system Hamiltonian of the form:
\begin{equation}
	\hat{ H }_{\rm pa}  =  - \Lambda \left( \hat{a}^2 + \hat{a}^{\dagger 2} + \hat{b}^2 + \hat{b}^{\dagger 2} \right) .
	\label{eq:CompensationHamiltonian}
\end{equation}
This is written in the interaction picture; in the non-rotating laboratory frame, this corresponds to an equal-strength parametric modulation of each oscillator's spring constant at the frequency $2 \omega_m$.  Writing this in the tilde quadrature basis of interest, we have:
\begin{equation}
\hat{ H }_{\rm pa} = \Lambda \left( \tilde{P}^2_- + \tilde{P}^2_+ - \tilde{X}^2_+ - \tilde{X}^2_- \right) . \label{eq:compensation}
\end{equation}
We thus see that by taking $\Lambda = p \tilde{\Omega} / 2$, these terms combine with the last term in Eq.~(\ref{eq:Hamtilded}) to give a Hamiltonian which
has the general back-action-evading form:  
\begin{eqnarray}
\hat{\mathcal{H}}_C =  \hat{\mathcal{H}} + \hat{ H }_{\rm pa} & = & \tilde{\Omega} \left( \tilde{X}_+ \tilde{X}_- + \tilde{P}_+ \tilde{P}_- \right) + \tilde{G} \tilde{X}_+ \hat{X}_c \nonumber \\
	& & + p\tilde{\Omega} \left( \tilde{P}^2_-  - \tilde{X}^2_+ \right) + \hat{H}_{\rm diss} .
	\label{eq:HamtildedComp} 
\end{eqnarray}
Note that the required value of the parametric modulation corresponds to requiring the amplitude of the spring constant modulation of each oscillator to satisfy $\Delta k_i/k_i = 4p\tilde{\Omega} / \omega_i$ ($i, =a,b$). In this Hamiltonian, the measured observables $\tilde{X}_+$ and $\tilde{P}_-$ are dynamically independent of their conjugate quadratures as required. The extra terms in the last line of Eq.~(\ref{eq:HamtildedComp}) only modify the dynamics of the unmeasured subsystem. We thus see that two-mode BAE is possible {\it even without} having perfectly symmetric optomechanical couplings. 

\section{Noise spectra}
\label{back-action}

To gain a better appreciation of the two-mode BAE scheme introduced above, we calculate the noise spectra of the measured and perturbed collective mechanical quadratures, and of the homodyne current from measurement of the cavity output. These may be calculated from the Heisenberg-Langevin equations in the usual manner \cite{TMSS}; the details are given in App.~\ref{SpectraCalculation}. Note that all quantum noise spectra quoted throughout this paper are \emph{symmetrized} quantum noise spectra.  Of particular interest here will be the effects of asymmetries (either in the optomechanical couplings or in the mechanical damping rates), which will cause deviations from the perfectly 
back-action-evading result. Readers not interested in the effects of imperfections (i.e. asymmetries) may wish to read Sec.~\ref{sec:idealcase} and then jump to Sec.~\ref{sec:forcesensitivity}, where we discuss beating the quantum limit on force sensing. 

\subsection{Ideal symmetric case}
\label{sec:idealcase}
We start with the ideal case of symmetric optomechanical couplings ($g_a = g_b$) and symmetric mechanical damping ($\gamma_a = \gamma_b = \gamma$).  The symmetrized noise spectral density of an observable $\hat{Z} \equiv [ \vec{V} ]_i $ is defined by
\begin{equation}
S_{Z} [ \omega ] \equiv \frac{1}{2} \int^{+\infty}_{-\infty} dt \ e^{i\omega t} \left\langle \left\{ \hat{Z} (t) , \hat{Z} (0) \right\} \right\rangle . \label{eq:symmnoise}
\end{equation}
For the measured observable $\hat{X}_+$ it is:
\begin{eqnarray}
S_{X_+}[\omega ] & = & \frac{1}{2} \left( \bar{n}_a + \bar{n}_b + 1 \right) \left[ \frac{\gamma /2}{ (\gamma /2)^2 + (\omega - \Omega )^2 } \right. \nonumber \\
& & \ \ \ \left. + \frac{\gamma/2}{(\gamma/2)^2 + (\omega + \Omega )^2} \right] , \label{eq:measuredobservable} 
\end{eqnarray}
where $\bar{n}_a$ and $\bar{n}_b$ are the temperatures of the two mechanical baths (expressed as a number of thermal quanta). We are working here in the interaction picture as always (i.e. $\omega$ is measured with respect to the average mechanical frequency $\omega_m$), and have assumed the good cavity (or ``resolved-sideband'') limit $\kappa \ll \omega_m$.  The spectrum of Eq.~(\ref{eq:measuredobservable}) consists of two Lorentzians, centred at $\pm \Omega$, and weighted by the thermal and quantum fluctuations of the mechanical oscillators. Since we are in the perfect back-action-evasion limit, there is no dependence on the optomechanical coupling $G$.  Turning to the perturbed observable $\hat{P}_+$, the spectrum is given, in the limit $\kappa \gg \Omega$, by:
\begin{eqnarray}
S_{P_+} [ \omega ] & = & \frac{1}{2} \left[ \bar{n}_a + \bar{n}_b + 1 + C ( 2\bar{n}_c + 1 ) \right] \nonumber \\
& & \ \ \ \times  \left[ \frac{\gamma /2}{ (\gamma /2)^2 + (\omega - \Omega )^2 } \right. \nonumber \\
& & \ \ \ \ \ \ \left. +  \frac{\gamma/2}{(\gamma/2)^2 + (\omega + \Omega )^2} \right]  ,  \label{eq:perturbedobservable}
\end{eqnarray}
where $\bar{n}_c$ is the number of thermal quanta associated with cavity dissipation, and we have quantified the optomechanical coupling $G$ in terms of the cooperativity parameter \cite{kimble:cooperativity}, defined as
\begin{equation}
C = \frac{2G^2}{ \gamma \kappa } . \label{eq:cooperativity}
\end{equation}
The cooperativity here is essentially the measurement strength, and thus describes the back-action heating rate of the perturbed collective quadrature.  It is equivalent to the quantity $n_{\rm BA}$ used in Ref.~\onlinecite{clerk:QND} to denote back-action heating of the perturbed quadrature (as a number of quanta) in optomechanical \emph{single-mode} back-action-evasion. 

We now turn to the spectrum of the cavity output.  We consider the case of a single-sided cavity, with homodyne detection of the cavity  output field.  The \emph{measured} spectrum of $\hat{X}_+$ is then defined in terms of the total homodyne current noise spectrum $S_I[\omega]$ as:
\begin{equation}
	S^{\rm meas}_{X_+}[ \omega ] = S_I[\omega ]/\mathcal{K}^2, \label{eq:homodynenoise}
\end{equation}
where $\mathcal{K} \equiv 2B G / \sqrt{\kappa/2} $ is a measurement gain with $B$ denoting the amplitude of the local oscillator in the homodyne detection scheme. Further details of the calculation of this quantity are given in App.~\ref{SpectraCalculation}.
Taking the most interesting limit $\kappa \gtrsim \Omega, \gamma$ (i.e.~the cavity is fast enough to measure the collective quadrature), 
 we find 
\begin{equation} 
	S^{\rm meas}_{X_+}[ \omega ] = S_{X_+ }[\omega ] + \frac{1/\gamma}{8C } (2\bar{n}_c + 1 ) . \label{eq:measuredspectrum}
\end{equation}
The second term here is the imprecision noise of the measurement (i.e.~output shot noise). In the perfect back-action-evading limit considered here, it can be made arbitrarily small without any resulting back-action penalty by increasing the cooperativity via the driving strength $\mathcal{E}_\pm$. Note that the standard quantum limit result at resonance (applicable when one is limited by back-action) is that the added noise must be at least as large as the zero-point motion of the oscillator.  We thus see that one beats this standard quantum limit constraint when 
\begin{equation}
C > (1 + 2 \bar{n}_c)/8. \label{eq:CforSQL}
\end{equation}
Typically the thermal occupation of the cavity, $\bar{n}_c$, is small, such that the constraint of Eq.~(\ref{eq:CforSQL}) is not experimentally demanding. Indeed, henceforth, we shall set $\bar{n}_c = 0$. The results accounting for non-zero $\bar{n}_c$ are readily obtained by multiplying cavity-dependent contributions to the noise spectra by $(2\bar{n}_c + 1)$. 

\subsection{Asymmetric optomechanical couplings and mechanical damping}
\label{subsec:AsymmetricNoise}

In the presence of coupling asymmetry, the measured collective quadrature is now $\tilde{X}_+$ (rather than $\hat{X}_+$) as defined in Eq.~(\ref{eq:tilded0}), and so it is the noise spectrum of this quantity that we seek.  Our goal is to understand whether coupling and damping asymmetries can be tolerated, or even exploited, for the purpose of force sensing beyond the usual quantum limits.  


In the presence of asymmetry, the two sets of collective quadratures $(\tilde{X}_+, \tilde{P}_-)$ and $(\tilde{X}_-, \tilde{P}_+)$ are coupled to one another, increasing the complexity of the spectra. In general, we write the noise spectral density as the sum of a contribution from thermal and quantum fluctuations of the mechanical oscillators, a contribution from the back-action of the cavity detector, and an imprecision noise contribution, as:
\begin{eqnarray}
S^{\rm meas}_{\tilde{X}_+ } [ \omega ] & = &  
S^{\rm th}_{\tilde{X}_+}[\omega ] + S^{\rm ba}_{\tilde{X}_+} [ \omega ] + S^{\rm imp}_{\tilde{X}_+} [ \omega ] . \label{eq:thermaladded}
\end{eqnarray}
Now in the fully symmetric case considered previously, we have $S^{\rm th}_{\tilde{X}_+}[\omega ] = S_{X_+}[\omega ]$ as given in Eq.~(\ref{eq:measuredobservable}), $S^{\rm ba}_{\tilde{X}_+}[\omega] = 0$ (since the measurement is BAE), and $S^{\rm meas}_{\tilde{X}_+}[\omega ]$ is given by Eq.~(\ref{eq:measuredspectrum}). 

Now the simplest contribution to the spectrum in the asymmetric case is the imprecision noise contribution. This is given by, c.f. Eq.~(\ref{eq:measuredspectrum}),
\begin{equation}
S^{\rm imp}_{\tilde{X}_+}[ \omega ] = \frac{1/\gamma}{8\tilde{C}} , 
\end{equation}
where $\gamma$ is now the \emph{average} mechanical damping rate, 
\begin{equation}
\gamma = (\gamma_a + \gamma_b) / 2 , \label{eq:avedamping}
\end{equation}
and we have introduced the \emph{rotated} cooperativity parameter by
\begin{eqnarray}
\tilde{C} = \frac{2\tilde{G}^2}{\gamma \kappa} 
& = & C \left[ 1 + (G_d/G)^2 \right] \nonumber \\
& = & C \sec^2 \left( \frac{\arctan p}{2} \right) , \label{eq:rotatedC}
\end{eqnarray}
which follows from Eqs.~(\ref{eq:pDefn}), (\ref{eq:GTilde}) and (\ref{eq:cooperativity}). 

The thermal and back-action contributions to the noise spectrum exhibit complicated dependencies on system asymmetries. They are given, in a general form, by Eqs.~(\ref{eq:thermalnoisespectraldensity}) and (\ref{eq:BAbit}), respectively. For weak asymmetries, however, one finds that the measured noise spectrum of $\tilde{X}_+$ still has approximately Lorentzian peaks at $\omega = \pm \Omega$ (not $\tilde{\Omega}$), as in the symmetric case.  The system asymmetries will be quantified through the dimensionless parameters $p$ and $d$, which quantify the degree of coupling and damping asymmetry respectively. The parameter $p$ is defined in Eq.~(\ref{eq:pDefn}), while $d$ is defined by:
\begin{equation}
	d  =  (\gamma_a - \gamma_b)/(\gamma_a + \gamma_b) . \label{eq:dDefn}
\end{equation}

Since the general expressions for these noise contributions are complicated, we shall focus on two special cases: the noise \emph{at} the effective mechanical resonance ($\Omega$ where $\Omega \gg \gamma$) and the noise \emph{far-detuned from} the effective mechanical resonance ($\Omega + \Delta$ where $\Omega \gg \Delta \gg \gamma$). Furthermore, we shall focus on the limit $\gamma/\Omega \rightarrow 0$. First we consider the system {\it in the absence} of the compensating parametric driving, such that the Hamiltonian is given by Eq.~(\ref{eq:HamUntilded}). In Sec.~\ref{sec:compensatedspectra} we shall consider the compensated system, with Hamiltonian given by Eq.~(\ref{eq:HamtildedComp}). 

\subsubsection{Noise at mechanical resonance}
\label{sec:noiseatmechanicalresonance}
First we consider the noise contributions at the effective mechanical resonance frequency. The noise contribution from thermal and quantum fluctuations is a complicated function of the coupling asymmetry, as this asymmetry modifies the effective mechanical susceptibility. 
However, in the case where the coupling asymmetry is zero ($p=0$), we simply have
\begin{eqnarray}
S^{\rm th}_{ X_+} [\Omega ] 
& = & \frac{1}{\gamma_a} \left( 1/2 + \bar{n}_a \right) + \frac{1}{\gamma_b} \left( 1/2 + \bar{n}_b \right) . \label{eq:thermalnoiseonresonance}
\end{eqnarray}
That is, the noise contribution from each mechanical oscillator is simply given by the thermal and quantum fluctuations of the bath to which it is coupled, scaled by its damping rate. 
Another case of interest, for reasons that shall become clear later, is that of \emph{matched} asymmetries ($p=-d$), and in this case we find
\begin{equation}
S^{\rm th}_{\tilde{X}_+} [ \Omega ] =  \frac{1}{\gamma} \left[ \left( 1/2 + \bar{n}_a \right) (1+d^3) + \left( 1/2 + \bar{n}_b \right) (1-d^3) \right] .
\end{equation}

\setlength{\abovecaptionskip}{-10pt }
\setlength{\belowcaptionskip}{-5pt }

\begin{figure}[th]
\begin{center}
\includegraphics[scale=0.48]{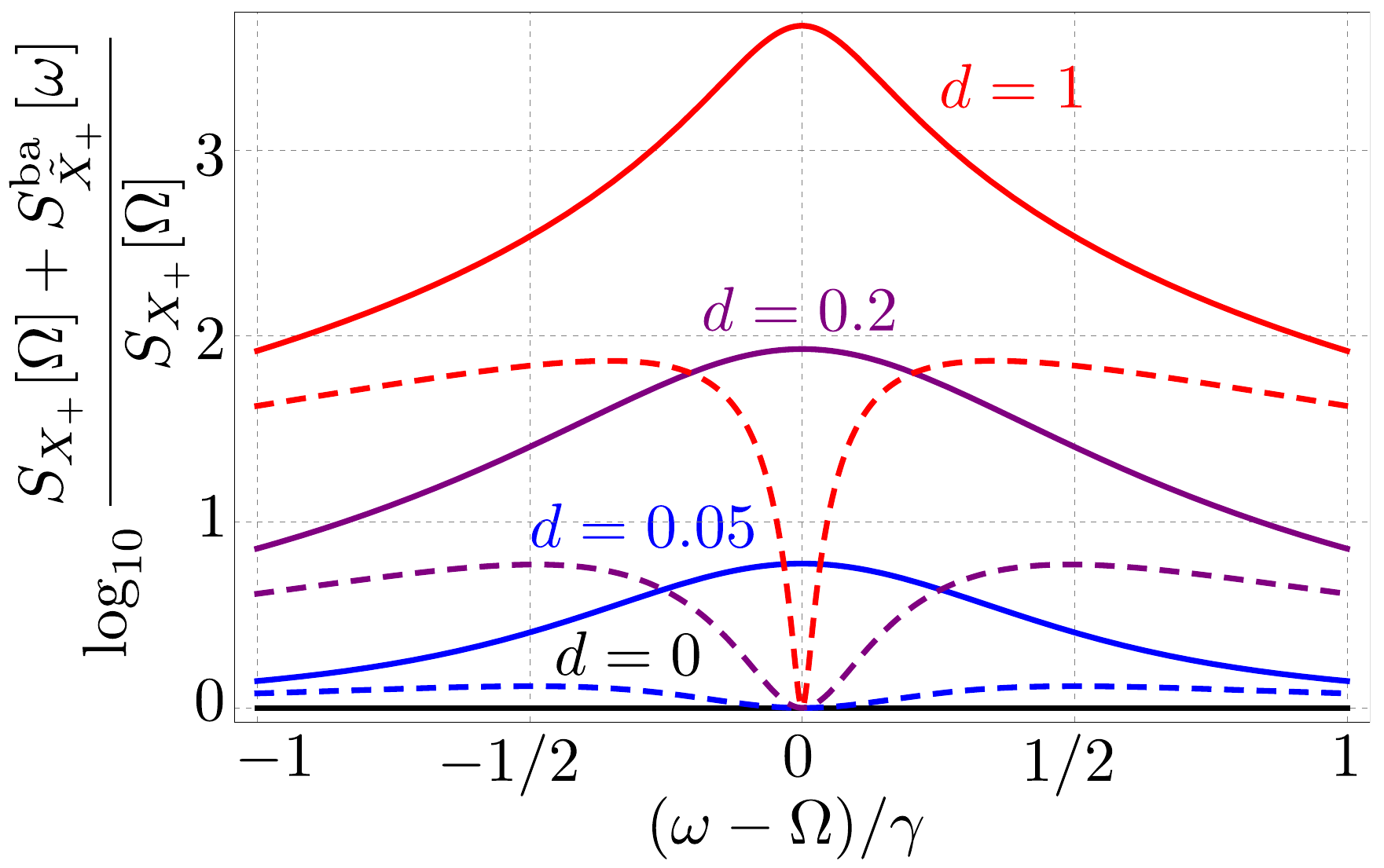}
\end{center}
\caption{ (Color online). Back-action noise, $S^{\rm ba}_{\tilde{X}_+}[\omega ]$, for a range of optomechanical coupling and mechanical damping asymmetries, centred on $\omega = +\Omega$. The plotted spectra are offset and normalised by the noise spectral density at $\Omega$ in the ideal, fully symmetric case, c.f. Eq.~(\ref{eq:measuredobservable}). In the fully symmetric case (black line), there is no back-action contribution to the spectrum. Asymmetries lead to back-action contributions to the spectra (solid lines). However, by matching asymmetries the back-action contribution on resonance may be nullified (dashed lines), see Eq.~(\ref{eq:perturbback-action}). The asymmetries for each line are (c.f. Eq.~(\ref{eq:dDefn})): $d=0, G_d/G=0$ (black); $d=0.05, G_d/G=0.025$ (blue, solid); $d=0.05, G_d/G=-0.025$ (blue, dashed); $d=0.2, G_d/G=0.1$ (purple, solid); $d=0.2, G_d/G=-0.1$ (purple, dashed); $d=1, G_d/G=0.414$ (red, solid); $d=1, G_d/G=-0.414$ (red, dashed). The other parameters used here (and in subsequent figures) are based on the experiments performed by Teufel and co-workers \cite{teufel:groundstate}, as described in Sec.~\ref{parameterlist}. }
\label{fig:Spectra}
\end{figure}

\setlength{\belowcaptionskip}{-10pt }


Next we consider the noise contribution due to the back-action of the cavity on the mechanical oscillators. At the resonant peaks, this is
\begin{equation}
S^{\rm ba}_{\tilde{X}_+} [\Omega ] = \frac{1}{\gamma} \frac{(p+d)^2(1+p^2)}{(1-d^2+p^2)^2} \tilde{C} . \label{eq:perturbback-action} 
\end{equation}
That is, the asymmetries lead to back-action heating which is proportional to the measurement strength $C$. As shown in Fig.~\ref{fig:Spectra}, in general, the back-action noise contribution increases in the presence of either coupling or damping asymmetries (as indicated by the solid lines). Surprisingly, however, by having appropriately tuned damping and coupling asymmetries such that $p+d = 0$, one can cancel this extra heating effect at resonance (as indicated by the dashed lines). The measured observable is driven by back-action from both the coupling asymmetry and damping asymmetry; at resonance these are precisely out-of-phase and may coherently cancel. Unfortunately, though, this cancellation occurs only in a small bandwidth about the mechanical resonance, as will be described quantitatively in Sec.~\ref{sec:forcesensitivity}. 
Recall that this back-action noise cancellation occurs in the limit $\gamma/\Omega \rightarrow 0$. To higher-order in $\gamma/\Omega$, and in the case of matched coupling and damping asymmetries ($p =-d$) we have:
\begin{eqnarray}
S^{\rm ba}_{ \tilde{X}_+} [ \Omega ] & = & \frac{1}{\gamma} \left( \frac{\gamma}{\Omega} \right)^2 \frac{1}{8} \nonumber \\
& & \times (1+d^2)^3 \left( 1+ d^2 - \sqrt{1+d^2} \right) C . \label{eq:spectraldensityasymmetric}
\end{eqnarray}
This result will be useful in determining the ultimate limit to force sensing in this system. 

\subsubsection{Noise away from mechanical resonance}
Next we consider the noise spectrum far from the effective resonance frequency ($\omega = \pm \Omega$), at a detuning $\Delta$ where $\Omega \gg \Delta \gg \gamma$. The thermal contribution in this limit is
\begin{eqnarray}
S^{\rm th}_{\tilde{X}_+} [ \Omega + \Delta ] & = & \frac{\gamma}{4\Delta^2} \left[ (1+d)  \left( \frac{1}{2} + \bar{n}_a \right) \left( 1 - \frac{p}{1+p^2} \right) \right. \nonumber \\ 
& & + \left. (1-d)  \left( \frac{1}{2} + \bar{n}_b \right) \left( 1 + \frac{p}{1+p^2} \right) \right] . \nonumber \\
& &  \label{eq:thermalnoiseoffresonance}
\end{eqnarray}
In the case of no asymmetries $(p,d=0)$, Eq.~(\ref{eq:thermalnoiseoffresonance}) reduces to a simple result consistent with Eq.~(\ref{eq:measuredobservable}). 

Next, the back-action contribution to the noise spectrum at $\omega = \pm (\Omega + \Delta)$ is 
\begin{equation}
S^{\rm ba}_{\tilde{X}_+}[\Omega + \Delta] = \frac{\gamma}{4\Delta^2} \frac{p^2}{1+p^2} \tilde{C} . \label{eq:baoffresonant}
\end{equation}
Clearly, far from the effective mechanical resonance, the back-action heating due to the presence of damping asymmetry 
is strongly suppressed compared to the heating due to coupling asymmetry. 
As we will see in Sec.~\ref{sec:forcesensitivity}, this will allow for excellent force sensing of a force applied to one of the mechanical oscillators which is strongly detuned from resonance.
In the absence of coupling asymmetry, the dominant back-action noise contribution is at higher-order in $\gamma/\Delta$,
\begin{equation}
S^{\rm ba}_{X_+} [\Omega + \Delta] = \frac{\gamma}{4 \Delta^2} \left( \frac{\gamma}{2\Delta} \right)^2 d^2 C . \label{eq:spectraldensityasymmetricdetuned}
\end{equation}  
That is, the back-action heating due to damping asymmetry is suppressed by the small factor $(\gamma/2\Delta )^2$.   

\subsubsection{Noise with compensation}
\label{sec:compensatedspectra}
Now we turn to the noise spectra \emph{in the presence} of the compensation described in Eq.~(\ref{eq:HamtildedComp}). The question is whether or not such compensation is useful for the purpose of force sensing, and to answer this question, we calculate the measured noise spectra. For weak asymmetries, the noise spectrum remains well-approximated by Eq.~(\ref{eq:measuredobservable}) if we make the replacement $\Omega \rightarrow \tilde{\Omega}$. That is, the effective collective mechanical oscillator frequency is now $\tilde{\Omega}$ (rather than $\Omega$), but again we shall quote results in two cases: at resonance ($\tilde{\Omega}$, assuming $\tilde{\Omega} \gg \gamma$), and far-detuned from resonance ($\tilde{\Omega}+\Delta$, assuming $\tilde{\Omega} \gg \Delta \gg \gamma$). 

At resonance, the thermal noise contribution is 
\begin{eqnarray}
S^{\rm th}_{\tilde{X}_+} [ \tilde{\Omega} ] & = & \frac{1}{ \gamma \left[ 1 - d^2 (1-p^2) \right] } \left[ \bar{n}_a + \bar{n}_b + 1 \right. \nonumber \\
& & \left. + (\bar{n}_b - \bar{n}_a ) d \frac{1 - d^2 (1+p^2)}{1-d^2(1-p^2)} \right] .
\end{eqnarray}
This reduces to the result of Eq.~(\ref{eq:thermalnoiseonresonance}) in the case $p=0$ (where $\tilde{\Omega} \rightarrow \Omega$). 
The back-action contribution is
\begin{equation}
S^{\rm ba}_{ \tilde{X}_+ } [ \tilde{\Omega} ] = \frac{1}{\gamma} \frac{d^2}{ \left[ 1 - d^2 (1-p^2) \right]^2 } \tilde{C} .  \label{eq:compensatedperturbative}
\end{equation}
Clearly, the back-action heating due to coupling asymmetry is now attenuated compared with the uncompensated case, c.f. Eq.~(\ref{eq:perturbback-action}).


Far from the effective mechanical resonance, the thermal noise contribution, $S^{\rm th}_{\tilde{X}_+} [ \tilde{\Omega} + \Delta ]$, is given by Eq.~(\ref{eq:thermalnoiseoffresonance}) with $p=0$, and the back-action noise contribution, $S^{\rm ba}_{\tilde{X}_{+}} [ \tilde{\Omega} + \Delta ]$, is given by Eq.~(\ref{eq:spectraldensityasymmetricdetuned}) provided that we make the replacement $C \rightarrow \tilde{C}$. In both cases, the noise contribution due to coupling asymmetry is suppressed by the use of compensation.

\subsection{Summary}

The noise spectrum of the measured observable may be expressed as a sum of contributions from thermal and quantum fluctuations, back-action noise and imprecision noise, as per Eq.~(\ref{eq:thermaladded}). In the fully symmetric case there is no back-action contribution, and the noise spectrum is given by Eqs.~(\ref{eq:measuredobservable}) and (\ref{eq:measuredspectrum}). In the asymmetric case, the back-action noise contribution is a complicated function of the asymmetries. At resonance it is possible to cancel this contribution by matching asymmetries (see Eq.~(\ref{eq:perturbback-action})), while the contribution far from resonance is relatively insensitive to damping asymmetry (see Eq.~(\ref{eq:spectraldensityasymmetricdetuned})). 
 
\section{Force Sensitivity}
\label{sec:forcesensitivity}

A key motivation for the two-mode BAE scheme is the possibility of continuously monitoring 
both quadratures of a narrow-band force without any back-action-related quantum limit.  We now analyze this
possibility in detail, paying close attention to the role of mechanical dissipation, something that was not discussed in Ref.~\onlinecite{tsang:evadeQM}.  

\subsection{Conventional quantum limits}

We consider the standard situation where a narrow-band force $F(t)$ is applied to one of the two mechanical oscillators in our setup (say the $a$ mechanical oscillator):
\begin{equation}
\hat{H}_F =F(t) \hat{x}_a  = f(t) \left( \hat{a} + \hat{a}^\dagger \right) , \label{eq:DrivingLabFrame}
\end{equation}
where $f (t) = \Delta x_a F(t)$. In the standard way, we will express the total noise in the measured homodyne signal as an equivalent added thermal noise on the driven oscillator. The measured force noise spectral density has contributions both from the added noise of the measurement as well as the inherent quantum (zero-point) and thermal fluctuations of the measured system, and takes the general form:
\begin{equation}
	S^{\rm meas}_{F} [\omega ] \equiv  \hbar m \gamma_a \omega \left( 1 + 2\bar{n}_a [\omega ] + 2\bar{n}_{\rm add} [\omega ] \right) , \label{eq:standardform}
\end{equation}
where $\bar{n}_a [\omega ]$ describes the thermal occupation of the driven oscillator and  $\bar{n}_{\rm add} [\omega ]$ is the noise added by the measurement. 
The most straightforward way to monitor $F(t)$ would be to continuously monitor the position $\hat{x}_a$ of the driven mechanical oscillator.  In this case, the full quantum limit 
on continuous position detection \cite{clerk:review} directly leads to a quantum limit on force detection:
\begin{equation}
	\bar{n}_{\rm add}[\omega] \geq 1/2 . 	\label{eq:ConventionalQL}
\end{equation}
We will refer to this as the \emph{full} quantum limit on continuous force detection:  the added noise at each frequency is at best equal to the zero-point noise.

Reaching the above-defined conventional quantum limit at frequencies far from the mechanical resonance requires strong correlations between the position
detector's back-action and imprecision noises \cite{clerk:review}, something that can be difficult to achieve.  In the absence of such correlations, 
one is subject to a more severe constraint, the so-called ``standard quantum limit".  For frequencies $\omega = \omega_a + \Delta$ far from the mechanical resonance
(i.e. $|\Delta| \gg \gamma_a$), this standard quantum limit takes the form:
\begin{equation}
	\bar{n}_{\rm add}[\omega_a + \Delta] \geq \Delta/\gamma_a .
	\label{eq:StandardQL}
\end{equation}

By now calculating the added noise $\bar{n}_{\rm add} [\omega ]$ of the two-mode BAE scheme which is the focus of this paper,  we can determine whether one can surpass these conventional quantum limits.  Naively, one might think that as one is evading the measurement back-action, there should be no quantum limit on the added noise.  The situation is, however, more complex.  The auxiliary oscillator in 
our scheme (mechanical oscillator $b$) will have its own quantum and thermal fluctuations which contribute to the output noise and thus the added noise of the measurement.  One cannot
simply set this to zero by making $\gamma_b = 0$, as if $\gamma_a \neq \gamma_b$, the complete BAE nature of the scheme is lost (as shown above).  Despite these complications, we
find that there are ways to perform force sensing using our scheme that yield added noise numbers far below the conventional quantum limits.

\subsection{Results for two-mode BAE}

The two-mode BAE scheme is capable of measuring both quadratures of the signal force $F(t)$ as long as it is contained in a 
 bandwidth $\ll \Omega$ about the mechanical resonance frequency $\omega_a$; we thus restrict attention to this case.  The linearized equations of motion readily yield
 the linear-response relation between the measured observable $\tilde{X}_+$ and the frequency components of the applied force (see App.~\ref{sec:TFs}).  For frequencies 
 $\delta$ satisfying $| \delta | \ll \Omega$, we find
\begin{equation}
	\langle \tilde{X}_+[\pm (\omega_2+\delta)] \rangle = \Delta x_a \ \chi_F[\pm (\omega_2+\delta)] F [ \pm (\omega_1 + \delta) ],
	\label{eq:ForceSusceptibility}
\end{equation}
where $\omega_2 = \Omega $ and $\omega_1 = \omega_a $ in the uncompensated case, and $\chi_F[\omega]$ is a transfer function (which satisfies $\chi_F[-\omega] = \chi_F[\omega]^*$) whose form follows from the equations of motion; we discuss its properties further in what follows. Note that in the compensated case the relevant frequencies are shifted such that $\omega_2 = \tilde{\Omega}$ and $\omega_1 = \omega'_a = \omega_a + \Omega (1-\cos 2\theta )$ with $\theta$ as defined in Eq.~(\ref{eq:xiDefn}), and for this reason results in this section shall be expressed in terms of the general frequencies $\omega_1$ and $\omega_2$. We shall consider both the cases of a mechanically resonant signal force $(\delta = 0)$ and a mechanically non-resonant signal force $(\delta = \Delta \gg \gamma)$. We stress here that the left-hand-side of Eq.~(\ref{eq:ForceSusceptibility}) involves the frequency components of $\tilde{X}_+[\omega]$ in the rotating frame used to write the system Hamiltonian in Eqs.~(\ref{eq:HamUntilded}) and (\ref{eq:HamtildedComp}),  whereas the frequency components of $F[\omega]$ are in the (non-rotating) laboratory frame. 

Eq.~(\ref{eq:ForceSusceptibility}) tells us that the full spectral information of the signal force is reproduced in the dynamics of the collective mechanical quadrature $\tilde{X}_+$, the only proviso being that the mechanical resonance frequency $\omega_1$ is effectively shifted to $\omega_2$.  It follows that the measured force noise spectral density is given by:  
\begin{eqnarray}
	S^{\rm meas}_{F} [ \omega_1 + \delta ] & = & \frac{\hbar^2 / ( \Delta x_a )^2  }{ |\chi_F [\omega_2+\delta]|^2} S^{\rm meas}_{\tilde{X}_+} [ \omega_2 + \delta ] ,\label{eq:forcenoisetransfer} 
\end{eqnarray}
where the calculation of the output noise $S^{\rm meas}_{ \tilde{X}_+} [\omega_2 + \delta ] $ was discussed in Sec.~\ref{back-action}; recall that these noise spectra were explicitly symmetric in frequency.

Writing the measured force noise spectral density in the form of Eq.~(\ref{eq:standardform}), we can express the added noise of our scheme as
\begin{eqnarray}
\bar{n}_{\rm add}[\omega] & = & \bar{n}^{\rm aux}_{\rm add} [\omega]+ \bar{n}^{\rm cav}_{\rm add}[\omega] \nonumber \\
& = & \bar{n}^{\rm aux}_{\rm add}[\omega] +  \bar{n}^{\rm ba}_{\rm add}[\omega] + \bar{n}^{\rm imp}_{\rm add}[\omega], \label{addednoisedefine}
\end{eqnarray}
where $\bar{n}^{\rm aux}_{\rm add}[\omega ]$ is the added noise associated with the inherent thermal and quantum fluctuations of the \emph{auxiliary} mechanical oscillator, and $\bar{n}^{\rm cav}_{\rm add}[\omega ]$ is the added noise associated with the coupling of the mechanical oscillators to the cavity. This second contribution may be further decomposed into back-action noise, $ \bar{n}^{\rm ba}_{\rm add}[\omega ]  $, and imprecision noise, $\bar{n}^{\rm imp}_{\rm add}[\omega ] $, contributions. These quantities follow from Eqs.~(\ref{eq:standardform}) and (\ref{eq:forcenoisetransfer}). 


While full expressions for the frequency-dependent added noise $\bar{n}_{\rm add}[\omega]$ may be readily derived, for simplicity we focus on the two main cases of interest:  
that of a mechanically-resonant signal force, and that of a far-detuned mechanical signal force. Henceforth in this section we consider results in the zero-temperature limit $( \bar{n}_a,\bar{n}_b = 0)$, allowing the possibility of force sensing near and beyond quantum limits. 

\subsubsection{Detection of a mechanically resonant force}

The resonant signal force corresponds to having $F(t)$ centred in a narrow bandwidth $\ll \gamma_a$ about the mechanical resonance frequency $\omega_1$. We can consider resonant force detection both without and with the compensation scheme described in Sec.~\ref{subsec:AsymmetricNoise}. From Eq.~(\ref{eq:ForceSusceptibility}), we see that we need to know the transfer function $\chi_F$ at the effective resonance frequency $\omega_2$ ($\delta = 0$).  One finds, in the regime $\omega_2 \gg \gamma$, that:
\begin{eqnarray}
	\chi_F [ \pm \omega_2] & = & \mp  \frac{i}{\gamma} g_{r}(p,d) , \label{eq:sensingsusceptibility}
\end{eqnarray}
where $g_{r}$ describes the modification of the susceptibility due to asymmetries; its full form as a function of the asymmetry parameters $p$ and $d$ is given by Eq.~(\ref{eq:gar}) or (\ref{eq:gcr}) in the uncompensated or compensated cases, respectively.

In the perfectly symmetric case (i.e.~both mechanical oscillators have identical optomechanical couplings and damping rates), $g_{r}=1$, implying that $\hat{X}_+$ responds to the applied force analogously to an oscillator driven on resonance. In this case,  from Eq.~(\ref{eq:measuredspectrum}) we find that the added noise of the force sensing scheme on resonance is 
\begin{equation}
\bar{n}_{\rm add} [ \omega_a ] = \frac{1}{2} + \frac{1}{8C} \rightarrow \frac{1}{2}.
\end{equation}
As expected, there is no back-action contribution to the added noise, and the added noise is a monotonically decreasing function of the measurement strength $C$.  Nonetheless, one cannot beat the standard quantum limit in this case. Even though the measurement back-action goes to zero, the added noise is still limited by the zero-point fluctuations of the auxiliary mechanical oscillator.  

Now we consider force sensing in the presence of asymmetries. First we consider the case of damping asymmetry without coupling asymmetry $(d\neq 0,p=0 )$. In this case the added noise due to the auxiliary mechanical oscillator is simply
\begin{equation}
\bar{n}^{\rm aux}_{\rm add} [ \omega_a ] = \frac{1}{2} \frac{1+d}{1-d} . \label{eq:auxnoise}
\end{equation}
This contribution may be understood in the following manner. As the damping rate of the auxiliary mechanical oscillator is increased beyond that of the driven oscillator (i.e. $d$ from $0 \rightarrow -1$), its noise spectrum is broadened. Therefore most of its noise is outside the force detection bandwidth of the driven oscillator, and the added noise of the force detection scheme \emph{at resonance} is reduced. If we consider the opposite limit, in which the damping rate of the auxiliary mechanical oscillator is reduced below that of the driven oscillator (i.e. $d$ from $0 \rightarrow 1$), the added noise contribution is increased. The force noise spectrum due to the auxiliary oscillator is narrowed, becoming sharply peaked about resonance. In either case, however, the damping asymmetry \emph{alone} will lead to a back-action noise contribution, as given by Eq.~(\ref{eq:perturbback-action}), that will typically dominate the noise due to the auxiliary oscillator. 

Next we consider the case of coupling asymmetry without damping asymmetry $(p\neq 0, d=0)$. Again, coupling asymmetry leads to back-action heating and therefore reduces force sensitivity. However, compensation, as per Eq.~(\ref{eq:HamtildedComp}), can be used to almost restore the system's force sensing capability. The added noise at the shifted mechanical resonance frequency, $\omega'_a$, is now
\begin{equation}
\bar{n}_{\rm add} [ \omega'_a ] = \frac{1}{2} \sec^2 \left( \frac{\arctan p}{2} \right) + \frac{1}{ 8 \tilde{C} } \rightarrow \frac{1}{2} ,
\end{equation}
with the limit (equal to the full quantum limit) being approached in the high-cooperativity, low coupling asymmetry regime. Again, the full quantum limit cannot be surpassed, however. 


The situation becomes more complex with \emph{both} damping and coupling asymmetries. In general, these asymmetries cause back-action heating and reduce the system's force sensitivity. Indeed, using Eq.~(\ref{eq:perturbback-action}), the added noise due to back-action, to second-order in $p$ and $d$, is
\begin{equation}
\bar{n}^{\rm ba}_{\rm add} [ \omega_a ] =   (p+d)^2 C . 
\end{equation}
The added noise now has terms which increase with the measurement strength $C$.

However, from Eq.~(\ref{eq:perturbback-action}), we know that it is possible to cancel back-action on resonance by matching asymmetries (i.e. setting $p=-d$). This is promising for force sensing, but we must also check the added noise due to the auxiliary mechanical oscillator. This contribution is
\begin{eqnarray}
\bar{n}^{\rm aux}_{\rm add} [ \omega_a ] & = & \frac{1}{2} \frac{1-d}{1+d} (1+d+d^2) \left( 1 + \frac{d}{\sqrt{1+d^2}}  \right) . \nonumber \\
& &  \label{eq:auxmatched}
\end{eqnarray}
Clearly, the added noise contribution of the auxiliary mechanical oscillator goes to zero when $d \rightarrow 1$, and combined with Eq.~(\ref{eq:perturbback-action}), this suggests the added noise of the scheme can go to zero with matched asymmetries. This suggestion is based on calculations performed in the limit $\gamma/\Omega \rightarrow 0$. We can check this possibility by calculating the added noise to higher-order in $\gamma/\Omega$. The added noise contribution due to back-action with matched asymmetries ($p=-d$) is
\begin{eqnarray}
\bar{n}^{\rm ba}_{\rm add} [ \omega_a ] & = & \left[ 1 - \sqrt{1+d^2} - d ( 1 - d - \sqrt{1-d^2} ) \right] \nonumber \\ 
& & \times \frac{(1+d^2)^3}{1+d}  \frac{1}{8} \left( \frac{\gamma}{\Omega} \right)^2 C . \label{eq:BAmatched}
\end{eqnarray}
From Eqs.~(\ref{eq:spectraldensityasymmetric}) and (\ref{eq:BAmatched}) we find, in the extreme damping asymmetry case ($d=1$ and $p=-1$), that the added noise due to the cavity is
\begin{equation}
\bar{n}^{\rm cav}_{\rm add} [\omega_a ] = \frac{\gamma^2}{2 \Omega^2}C +  \frac{3}{ 128 (\sqrt{2}-1) C} .
\end{equation}
While there is a back-action contribution here (first term), it is proportional to $(\gamma / \Omega)^2$ and hence small. One readily finds an optimal measurement strength, 
\begin{equation}
C_0 = \frac{ \sqrt{3} }{ 8\sqrt{ \sqrt{2} - 1 } } \frac{\Omega}{\gamma} , \label{eq:optimalmeasurement}
\end{equation}
at which point the total added noise is 
\begin{equation}
\left. \bar{n}_{\rm add}  [\omega_a ] \right|_{C_0} = \frac{\gamma}{\Omega} \frac{ \sqrt{3} }{ 8 \sqrt{ \sqrt{2} - 1 } } .
\end{equation}
The added noise remains close to zero provided that $\Omega \gg \gamma$, and so the full quantum limit can be beaten for resonant force sensing via \emph{matched asymmetries}. 

Unfortunately, this result applies over only a very small bandwidth about the mechanical resonance frequency. The added noise may be evaluated as a function of frequency, and we are interested in the bandwidth $B$ over which the added noise remains below the full quantum limit. It may be shown that, in the case of extreme damping asymmetry ($d \rightarrow 1$), this bandwidth is
\begin{equation}
B = \frac{1}{2\sqrt{2}} \frac{\gamma}{\sqrt{C}} \sim \gamma \sqrt{ \frac{\gamma}{\Omega} } , 
\end{equation}
with the second scaling attained for an optimal measurement strength (cooperativity),
see Eq.~(\ref{eq:optimalmeasurement}). That is, the useful bandwidth becomes very small.

\begin{figure}[th]
\begin{center}
\includegraphics[scale=0.37]{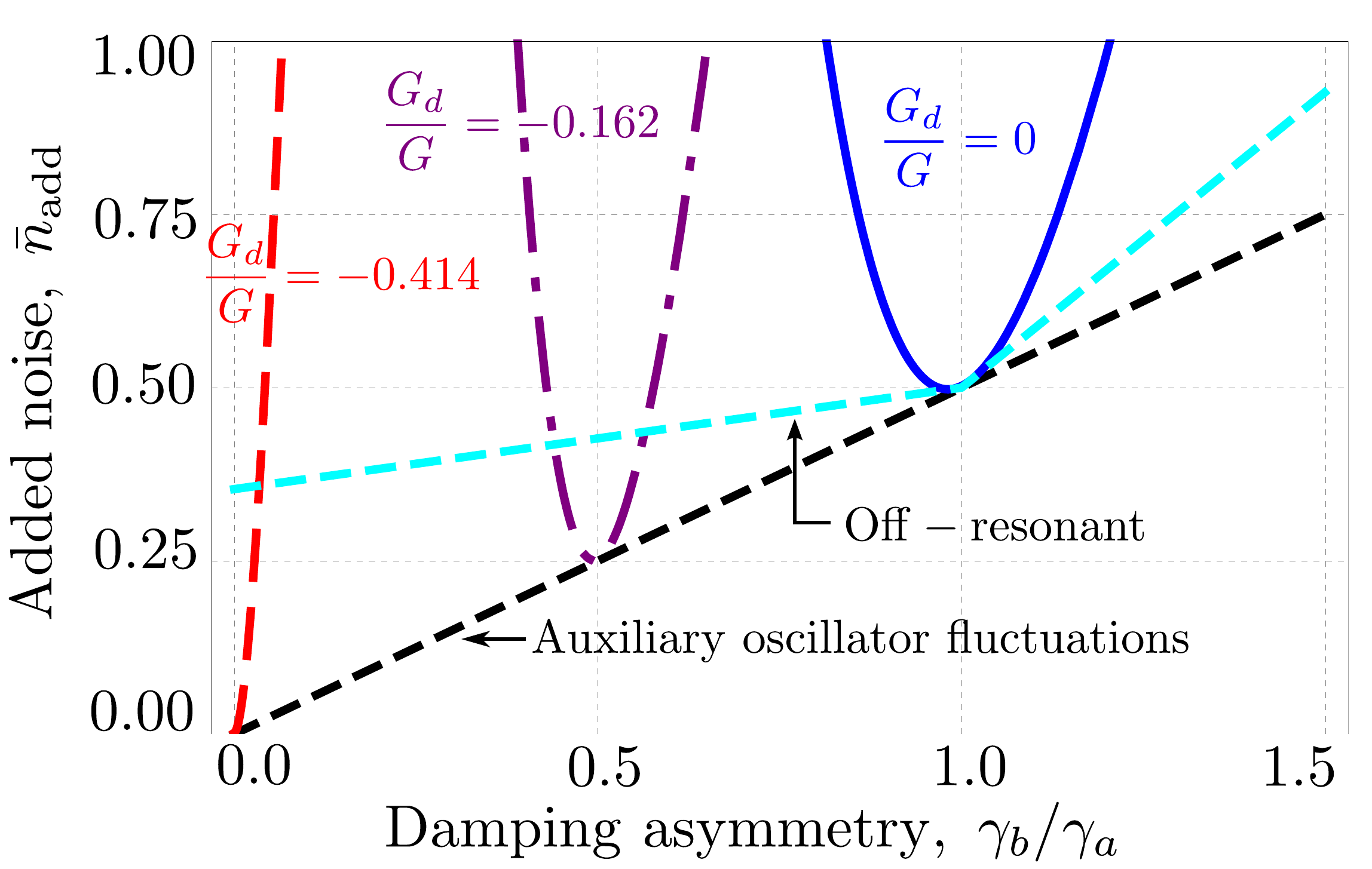}
\end{center}
\caption{ (Color online). Added noise for resonant ($\omega = \omega_a$) and non-resonant ($\omega = \omega_a + \Delta$) force sensing, $\bar{n}_{\rm add}[\omega]$, as a function of the damping rate of the auxiliary mechanical oscillator, $\gamma_b/\gamma_a$, while $\gamma_a$ is held fixed. The dashed black line is a lower bound on the contribution due to auxiliary oscillator thermal fluctuations; see Eqs.~(\ref{eq:auxmatched}) and (\ref{eq:AuxNoiseOffResonant}), for example. The blue, purple and red curves correspond to the added noise for resonant force sensing at $G_d/G = 0, -0.162$ and $-0.414$, respectively. The back-action contribution to the added noise can be made to go to zero on resonance by matching asymmetries, irrespective of the size of the damping asymmetry. The cyan line is the added noise for off-resonant force sensing, with no coupling asymmetry and the measurement strength optimised. } 
\label{fig:nadd}
\end{figure}

Plots of the added noise, as a function of the damping rate of the auxiliary oscillator (with the damping rate of the driven oscillator held constant), are shown in Fig.~\ref{fig:nadd}. The added noise is shown for three coupling asymmetries, and in each case the added noise at resonance may be reduced to the level of the auxiliary oscillator fluctuations by matching coupling and damping asymmetries. 

\subsubsection{Detection of a mechanically non-resonant force}

We now turn to the case where the signal force $F(t)$ is detuned from the mechanical resonance, and again we will consider both the original and compensated systems. Specifically, we consider the case where $F(t)$ 
is contained in a bandwidth $B$ centred on a frequency $\omega_1 + \Delta $, where $| \Delta | \gg \gamma_a$ and $B \leq | \Delta|$. From Eq.~(\ref{eq:ForceSusceptibility}), we see that we need to know the transfer function at $\omega_2 + \Delta$,
\begin{eqnarray}
	\chi_F [ \pm (\omega_2 + \Delta)] & = &
		  \frac{1}{2 \Delta} g_{n}(p,d) , \label{eq:modulationnonresonant}
\end{eqnarray}
where $g_{n}$ describes the modification of the transfer function due to asymmetries; its full form as a function of the asymmetry parameters $p$ and $d$ is given by Eq.~(\ref{eq:gao}) in the original case and by Eq.~(\ref{eq:gco}) in the compensated case. As compared with the case of a mechanically resonant signal force, see Eq.~(\ref{eq:sensingsusceptibility}), the symmetric ``gain'' is now $1/(2\Delta )$ rather than $1/\gamma$. Also note that force detection, in this case, is \emph{not} limited by a resonant detection bandwidth of order $ \gamma$. 

In the perfectly symmetric case (i.e.~both mechanical oscillators have identical optomechanical couplings and damping rates), $g_{n}=1$.  Then we find that the added noise is
\begin{equation}
	\bar{n}_{\rm add} [ \omega_a + \Delta ] = \frac{1}{2} + \left( \frac{2\Delta}{ \gamma } \right)^2  \frac{1}{8C} \rightarrow \frac{1}{2} , 
	\label{eq:NAddDetunedSymm}
\end{equation}
with the limit being approached in the high-cooperativity regime. Clearly, the scheme surpasses the standard quantum limit on non-resonant force detection (c.f. Eq.~(\ref{eq:StandardQL})).  It, however, is only
equal to the full quantum limit (c.f. Eq.~(\ref{eq:ConventionalQL})) that comes from allowing detector noise correlations.  Again, the residual added noise of $1/2$ a quantum in Eq.~(\ref{eq:NAddDetunedSymm}) is due to the zero-point noise of the auxiliary mechanical oscillator.  

We can easily determine the effects of asymmetries on our force sensing scheme. From Eq.~(\ref{eq:baoffresonant}), to second-order in the asymmetry parameters and in the limit $\gamma/\Delta \rightarrow 0$, we find
\begin{eqnarray}
\bar{n}^{\rm ba}_{\rm add} [\omega_a + \Delta ] & = & 
p^2 C . \label{eq:offresonantBAlimit}
\end{eqnarray}
As already noted, in this highly-detuned regime, it is only the optomechanical coupling asymmetry that yields an appreciable deviation from being quantum-limited.  This suggests that one could improve on the symmetric added noise result of Eq.~(\ref{eq:NAddDetunedSymm}) by exploiting damping asymmetry to reduce the added noise contribution from the auxiliary oscillator. 

Indeed, these added noise contributions due to the auxiliary mechanical oscillator are given by
\begin{subequations}
\begin{eqnarray}
\bar{n}^{\rm aux}_{\rm add} [ \omega_a + \Delta ] & = & \frac{1}{2} \frac{1-d}{1+d} \left[ 1 + p (p+\sqrt{1+p^2}) \right] , \nonumber \\
& & \label{eq:AuxNoiseOffResonant} \\
\bar{n}^{\rm aux}_{\rm add} [ \omega'_a + \Delta ] & = & 
\frac{1}{2} \frac{1-d}{1+d} \sec^2 \left( \frac{\arctan p}{2} \right) , \label{eq:AuxNoiseComp}
\end{eqnarray}
\end{subequations}
in the original and compensated cases, respectively. In both cases, they go to zero in the limit $d \rightarrow 1$. Combined with the insensitivity of the back-action noise to damping asymmetry as expressed in Eq.~(\ref{eq:offresonantBAlimit}), this would appear to enable force sensing beyond the full quantum limit, provided that there is no coupling asymmetry. In fact, we might expect that this force sensitivity can be achieved in the presence of coupling asymmetry if compensation is used. 

These arguments can be made precise. The following results describe the case \emph{with compensation included}, and so reduce to the case of no compensation when there is no coupling asymmetry to begin with ($p=0$ such that $\omega'_a \rightarrow \omega_a$). The added noise associated with the cavity is
\begin{eqnarray}
\bar{n}^{\rm cav}_{\rm add} [ \omega'_a + \Delta ] & = & \frac{\gamma^2}{4\Delta^2} \frac{d^2}{1+d} \sec^4 \left( \frac{\arctan p}{2} \right) C \nonumber \\
& & + \frac{4\Delta^2}{\gamma^2} \frac{1}{1+d} \frac{1}{8C} . 
\end{eqnarray}
This added noise is minimised by a cooperativity,
\begin{equation}
C_0 = \frac{4\Delta^2}{\gamma^2} \frac{1}{2\sqrt{2} |d|} \cos^2 \left( \frac{\arctan \ p}{2} \right) , 
\end{equation}
at which point the added noise due to the cavity is 
\begin{eqnarray}
\left. \bar{n}^{\rm cav}_{\rm add}[ \omega'_a + \Delta ]  \right|_{C_0} & = & \frac{1}{\sqrt{2}} \frac{ |d| }{|1+d|}\sec^2 \left( \frac{\arctan p}{2} \right). \nonumber \\ 
& & 
\end{eqnarray}
Taking the limit $d \rightarrow 1$, the added noise due to the auxiliary oscillator vanishes as per Eq.~(\ref{eq:AuxNoiseComp}), and the total added noise becomes
\begin{equation}
\left. \bar{n}_{\rm add}[ \omega'_a + \Delta ]  \right|_{C_0} = \frac{1}{2\sqrt{2}} \sec^2 \left( \frac{\arctan p}{2} \right) \rightarrow \frac{1}{2 \sqrt{2} } , \label{eq:addnoise22}
\end{equation}
with the result tending to the limit as the coupling asymmetry, $p \rightarrow 0$. Now, one can not only beat the standard quantum limit for non-resonant force detection (c.f. Eq.~(\ref{eq:StandardQL})) by a greater amount, but one can also surpass the full quantum limit on force detection (c.f. Eq.~(\ref{eq:ConventionalQL})). Without compensation, the added noise in Eq.~(\ref{eq:addnoise22}) grows more rapidly as $\left| p \right|$ increases.

\subsection{Summary}
In summary, for sensing of a mechanically-resonant force, force sensing \emph{at} the full quantum limit (being equal to the standard quantum limit, on resonance) is possible in the symmetric case. Even with coupling asymmetry, the possibility of force sensing near this quantum limit is retained via compensation. Further, it is possible to \emph{surpass} this full quantum limit via matched asymmetries, though only in a very narrow bandwidth. 

For sensing of a mechanically non-resonant force we can \emph{surpass} the standard quantum limit in the symmetric case. Furthermore, one may exploit damping asymmetry to surpass the full quantum limit. In both cases, these results can be retained in the presence of coupling asymmetry using compensation. Additionally, force sensing in this case is not limited to a resonant detection bandwidth. 



\section{Conditional Variances and Entanglement}
\label{sec:conditionalvariances}
In Sec.~\ref{back-action} we calculated the noise spectra of measured and perturbed observables in our system, with a view to analysing its force sensing capabilities in Sec.~\ref{sec:forcesensitivity}. These are all \emph{unconditional} quantities; that is, the description of the state of the system is not updated based on the measurements made upon it. However, since we are continuously monitoring the system, we can also discuss its \emph{conditional} dynamics, arising from updating the best estimate of the state of the system based on the measurement record. In particular, a continuous measurement conditionally projects the system into an eigenstate of the measured observable. This shall enable, in this system, conditional mechanical two-mode squeezing and the entanglement of the two mechanical oscillators. 

\subsection{All-Mechanical Entanglement}
The preparation and verification of macroscopic, all-mechanical entanglement is a fundamental goal of the study of mechanical systems in the quantum regime \cite{woolleyreview}. Such a state is also a physical approximation of an Einstein-Podolsky-Rosen channel \cite{EPR}, a key ingredient in quantum information processing protocols with continuous variables \cite{weedbrookRMP}. Electromechanical entanglement \cite{cleland:groundstate} and the entanglement of phonons in bulk \cite{walmsley:entanglediamond}, \emph{at the single-phonon level}, have both been demonstrated, though this is not the case for mechanical continuous variables. The entanglement of collective spin operators of atomic ensembles \cite{polzik}, and of the motional states of trapped ions \cite{wineland}, have both been achieved, however. 

There has been a great deal of discussion of the possibility of entangling macroscopic mechanical degrees of freedom with other degrees of freedom (see \cite{vitali1} and references therein), including other mechanical degrees of freedom. Particular attention has been paid to the possibility of all-mechanical entanglement with the interaction being mediated by an electromagnetic field. Proposals studied include placing mechanical oscillators in a ring cavity or interferometer, both without \cite{mancini,heidmann:doublecavity,chen} and with \cite{vitali:FBCoolCoupledMode} measurement and feedback, placing dielectric membranes in a cavity \cite{plenio}, and using remote optomechanical systems, both with \cite{lloyd,borkje,genes:remote} and without \cite{braunstein,joshi} protocols dependent on optical measurements.   

More recently, Schmidt and co-workers have proposed an all-mechanical entanglement generation scheme based on the detuned, modulated (and therefore, two-tone) driving of a coupled electromagnetic cavity \cite{marquardt:nanomechanicalCV}. Compared with their approach, our scheme has the considerable advantages that it does not require highly-detuned driving of the cavity, and is robust to the initial thermal populations of the mechanical oscillators. Further, the possibility of entanglement verification is built into our proposal. 

\subsection{Conditional Dynamics in Adiabatic Limit}
\label{adiabaticlimit}
The conditional dynamics of the system may be described in the standard manner using a stochastic master equation \cite{wiseman:measurecontrol}. The evolution of the joint density operator, $\sigma$, of the two mechanical modes and one electromagnetic mode, under homodyne detection of the cavity mode quadrature $\hat{c} e^{i\phi} + \hat{c}^\dagger e^{-i\phi}$ with a quantum efficiency $\eta$, is given by 
\begin{eqnarray}
d\sigma & = & - i [ \mathcal{\hat{H}}, \sigma ] \ dt + \kappa ( \bar{n}_c + 1 ) \mathcal{D}\left[ \hat{c} \right] \sigma \ dt \nonumber \\
& & + \kappa \bar{n}_c  \mathcal{D}\left[ \hat{c}^\dagger \right] \sigma \ dt + \sqrt{\eta \kappa } \mathcal{M} [ \hat{c} e^{i\phi} ] \sigma \ dW \nonumber \\
& & + \gamma_a ( \bar{n}_a + 1 ) \mathcal{D}[\hat{a}] \sigma \ dt + \gamma_a \bar{n}_a \mathcal{D} [\hat{a}^\dagger ] \sigma \ dt \nonumber \\
& & + \gamma_b (\bar{n}_b + 1 ) \mathcal{D}[\hat{b}] \sigma \ dt + \gamma_b \bar{n}_b \mathcal{D}[\hat{b}^\dagger ] \sigma \ dt , \label{eq:fullME}
\end{eqnarray}
where $\mathcal{\hat{H}}$ denotes the original or compensated Hamiltonian of Eq.~(\ref{eq:Hamtilded}) or (\ref{eq:HamtildedComp}), respectively, $\mathcal{D} [ \hat{A} ] \sigma \equiv \hat{A} \sigma \hat{A}^\dagger - \frac{1}{2} \hat{A}^\dagger \hat{A} \sigma - \frac{1}{2} \sigma \hat{A}^\dagger \hat{A}$ is the dissipative superoperator, $\mathcal{M} [ \hat{A} ] \sigma = \hat{A} \sigma + \sigma \hat{A}^\dagger - {\rm Tr} \ [ \hat{A} \sigma + \sigma \hat{A}^\dagger ] \sigma $ is the measurement superoperator, and $dW$ is the Wiener increment. 

We consider the ``good measurement'' limit, in which the cavity damping rate exceeds the rates in the Hamiltonians of Eq.~(\ref{eq:Hamtilded}) or (\ref{eq:HamtildedComp}); that is, $\kappa > \Omega, p\tilde{\Omega},\tilde{G}$ ($\tilde{\Omega}$ rather than $\Omega$ in the compensated case). In this limit, the cavity responds rapidly to the dynamics of the coupled mechanical oscillators, and we may adiabatically eliminate the cavity mode \cite{wiseman:adiabaticelim}. Setting $\phi = \pi/2$ for convenience, the cavity mode lowering operator is given by $\hat{c} =  - i \sqrt{2} \tilde{G} \tilde{X}_+ /\kappa$. Note that it is crucial that the adiabatic elimination is performed in terms of the rotated observables; otherwise, the dissipative terms in Eq.~(\ref{eq:fullME}) will involve linear combinations of mechanical annihilation operators, which greatly complicates the description. 

Now the evolution of the (reduced) joint density operator of the two mechanical oscillators having traced out the cavity mode, $\rho = {\rm Tr}_{\rm cav} \left[ \sigma \right] $, is given by
\begin{eqnarray}
d\rho & = & - i [ \mathcal{\hat{H}}' , \rho ] \ dt 
\nonumber \\
& & -\frac{\Gamma}{2} [ \tilde{X}_+, [ \tilde{X}_+, \rho ] ] \ dt  + \sqrt{\eta \Gamma} \mathcal{M} [ \tilde{X}_+ ] \rho \ dW \nonumber \\ 
& & + \gamma_a ( \bar{n}_a + 1 ) \mathcal{D}[\hat{a}] \rho \ dt + \gamma_a \bar{n}_a \mathcal{D} [\hat{a}^\dagger ] \rho \ dt \nonumber \\
& & + \gamma_b (\bar{n}_b + 1 ) \mathcal{D}[\hat{b}] \rho \ dt + \gamma_b \bar{n}_b \mathcal{D}[\hat{b}^\dagger ] \rho \ dt , \label{eq:eliminatedME}
\end{eqnarray}
where $\mathcal{\hat{H}}'$ now denotes either the Hamiltonian of Eq.~(\ref{eq:Hamtilded}) or (\ref{eq:HamtildedComp}), \emph{excluding terms involving the electromagnetic mode operators}. We have also introduced the collective quadrature measurement rate as 
\begin{equation}
\Gamma \equiv \gamma \tilde{C} = 2\tilde{G}^2/\kappa , \label{eq:measurementrate}
\end{equation}
where $\tilde{G}$ is the modified coupling constant introduced in Eq.~(\ref{eq:GTilde}).
The measurement record increment associated with Eq.~(\ref{eq:eliminatedME}) is given by
\begin{equation} 
dr = \langle \tilde{X}_+ \rangle dt + dW/\sqrt{4\eta \Gamma}. \label{eq:measurementrecordincrement}
\end{equation} 
Integrating Eq.~(\ref{eq:eliminatedME}) according to the measurement record increment of Eq.~(\ref{eq:measurementrecordincrement}) allows us to continually update our best estimate of the state of the two mechanical oscillators. 

\subsection{Best Estimates of Quadratures and Conditional Variances}
Now according to Eq.~(\ref{eq:eliminatedME}), with the quadratic Hamiltonian obtained from Eq.~(\ref{eq:Hamtilded}) or (\ref{eq:HamtildedComp}), and linear damping, decoherence and measurement, the steady-state reduced density operator $\rho$ will be Gaussian. Therefore, the conditional state of the system will be fully described by knowledge of its first and second moments. One can immediately write down the equations of motion for the (scalar) best estimates of the collective quadrature observables and for the conditional variances corresponding to the stochastic master equation in Eq.~(\ref{eq:eliminatedME}). The vector of best estimates is formed as
\begin{equation}
\vec{\bar{V}} = \left( \bar{X}_+, \bar{P}_-, \bar{X}_-, \bar{P}_+ \right)^T . \label{eq:bestestimatesvec}
\end{equation}
For later convenience, we introduce the following notation for the ten independent elements of the two-mode, symmetrically-ordered covariance matrix, $\mathbf{\Sigma}$:
\begin{equation}
\mathbf{\Sigma} = \left[ \begin{array}{cccc} V_{\tilde{X}_+} & \Sigma_{+-} & \Sigma_{XX} & \Sigma_{++} \\ \Sigma_{+-} & V_{\tilde{P}_-} & \Sigma_{--} & \Sigma_{PP} \\ \Sigma_{XX} & \Sigma_{--} & V_{\tilde{X}_-} & \Sigma_{-+} \\ \Sigma_{++} & \Sigma_{PP} & \Sigma_{-+} & V_{\tilde{P}_+}  \end{array} \right] \label{eq:CM} .
\end{equation}
In general, the elements of $\mathbf{\Sigma}$ will describe covariances of the rotated observables, though we drop the tilde notation in the subscripts of the off-diagonal covariances in Eq.~(\ref{eq:CM}) for convenience. In the symmetric case, the tilde notation on the subscripts of the collective quadrature variances in Eq.~(\ref{eq:CM}) shall be dropped. 

The equations for the best estimates of the collective (sum and difference) mechanical quadrature observables take the form of an Ornstein-Uhlenbeck process (c.f Eq.~(\ref{eq:QLE})), 
\begin{equation}
\frac{d}{dt} \vec{\bar{V}} = \mathbf{M} \cdot \vec{\bar{V}} + \vec{Q} \cdot \xi (t), \label{eq:bestestimates}
\end{equation}
where the system matrix $\mathbf{M}$ is given by Eq.~(\ref{eq:rotatedsystem}) or (\ref{eq:compensatedsystem}), for the Hamiltonian of Eq.~(\ref{eq:Hamtilded}) or (\ref{eq:HamtildedComp}), respectively, the measurement noise weighting vector is
\begin{equation}
\vec{Q} = \sqrt{4\eta \Gamma} \left( V_{\tilde{X}_+}, \Sigma_{+-}, \Sigma_{XX}, \Sigma_{++} \right)^T , \label{eq:measurementnoiseweight}
\end{equation}
and $\xi (t)=dW/dt$ is a white noise process describing the noise in the measurement. 

The equation for the best estimates of the covariances of the collective mechanical quadrature observables is
\begin{equation}
\dot{\mathbf{\Sigma}} = \mathbf{M} \mathbf{\Sigma} + \mathbf{\Sigma} \mathbf{M}^T + \mathbf{L} - \mathbf{\Sigma} \mathbf{K}^T \mathbf{K} \mathbf{\Sigma} , \label{eq:RiccatiEqn}
\end{equation}
where $\mathbf{\Sigma}$ is the symmetrically-ordered two-mode covariance matrix in the ordered basis defined by Eq.~(\ref{eq:rotatedvector}), $(\tilde{X}_+,\tilde{P}_-,\tilde{X}_-,\tilde{P}_+ )$, and again $\mathbf{M}$ is given by Eq.~(\ref{eq:rotatedsystem}) or (\ref{eq:compensatedsystem}). The matrices $\mathbf{K}$ and $\mathbf{L}$ are defined by
\begin{subequations}
\begin{eqnarray}
(\mathbf{K})_{11} & = & \sqrt{4\eta \Gamma}, \  \ (\mathbf{K})_{ij} = 0, \\
\mathbf{L} & = & \gamma \left[ \begin{array}{cccc}  \bar{n}'_{\rm eq} & 0 & \bar{n}'_{\rm d} & 0 \\  0 & \bar{n}'_{\rm eq} & 0 & \bar{n}'_{\rm d} \\ \bar{n}'_{\rm d} & 0 & \bar{n}'_{\rm eq}  & 0 \\ 0 & \bar{n}'_{\rm d} & 0 & \bar{n}'_{\rm eq} \end{array} \right],
\end{eqnarray}
\end{subequations}
where we have introduced the notation $\bar{n}'_{\rm eq} = \bar{n}_{\rm th} + 1/2 + d \bar{n}_{\rm d} $ and $ \bar{n}'_{\rm d} = \bar{n}_{\rm d} + d ( \bar{n}_{\rm th} + 1/2 )$, in terms of the dimensionless damping asymmetry introduced in Eq.~(\ref{eq:dDefn}) and the effective mechanical thermal occupations,  
\begin{subequations}
\begin{eqnarray}
\bar{n}_{\rm th} & = & \frac{1}{2} \left( \bar{n}_a + \bar{n}_b \right) , \label{eq:occupations1} \\
\bar{n}_{\rm d} & = & \frac{1}{2} \left( \bar{n}_a - \bar{n}_b \right) . \label{eq:occupations2}
\end{eqnarray}
\end{subequations}
Note that an equation of the form of Eq.~(\ref{eq:RiccatiEqn}) has previously been obtained in a study of the optimal control of two-mode entanglement (parametrically-coupled optical modes) under feedback \cite{mancini:optimalcontrol}.  

Due to the rules of the It\={o} calculus, Eq.~(\ref{eq:RiccatiEqn}) is a \emph{deterministic} system and we may solve for the steady-state covariance matrix. The steady-state equation corresponding to Eq.~(\ref{eq:RiccatiEqn}) is a continuous-time algebraic Riccati equation (CARE). This is a system of nonlinear algebraic equations, though one for which numerical methods are well-developed. 

\subsection{Steady-State Conditional Variances and Entanglement}
The two-mode, symmetrically-ordered covariance matrix $\mathbf{\Sigma}$ enables a full characterization of the entanglement and purity of the system \cite{CV}. Indeed, the entanglement may be directly quantified using the logarithmic negativity \cite{vidal:lognegativity}. From an experimental perspective, however, the estimation of all elements of the covariance matrix is challenging. A simpler method for determining whether or not the mechanical oscillators are entangled is provided by Duan's inseparability criterion \cite{duan:inseparability}. Note that this is a sufficient, but not necessary, condition for the inseparability of a bipartite quantum state. Specifically, we consider a generalized version of Duan's criterion \cite{giovannetti}, that here takes the form
\begin{equation}
V_{\tilde{X}_+ } + V_{ \tilde{P}_- } < \cos 2\theta = \frac{1-G^2_d/G^2}{1+G^2_d/G^2} . \label{eq:generalizedDuan}
\end{equation}
In the case of symmetric optomechanical coupling ($G_d=0$), the criterion of Eq.~(\ref{eq:generalizedDuan}) reduces to the standard version of Duan's criterion, 
\begin{equation}
V_{X_+} + V_{P_-} < 1 . \label{eq:Duan}
\end{equation}
Crucially, both quantities on the left-hand-side of Eq.~(\ref{eq:generalizedDuan}) may be obtained in a straightforward manner from the measurement record. The quantity $\tilde{X}_+$ is measured directly, while in the regime $\Omega \gg \gamma$, $\tilde{P}_-$ is dynamically coupled to it such that it is also effectively measured. 

\subsection{The Symmetric Case}

\subsubsection{Conditional Variances}
\label{subsec:ConditionalVariances}
In the ideal, symmetric case, an analytic form for the steady-state covariance matrix can be found (see App.~\ref{bestestimates}). Of particular interest is the variance of the measured collective quadrature. 
Taking the limit $\gamma/\Omega \rightarrow 0$ while the cooperativity $C$ is fixed leads to
\begin{equation}
V_{X_+} = \frac{ \sqrt{ 8\eta C ( \bar{n}_{\rm th} + 1/2 ) + 1 } - 1 }{4\eta C} \rightarrow \frac{ \sqrt{\bar{n}_{\rm th} + 1/2} }{\sqrt{2 \eta}} \frac{1}{\sqrt{C}} , \label{asymptoticVXp}
\end{equation}
with the latter result following for $C \gg 1/[8\eta ( \bar{n}_{\rm th} + 1/2)]$. This is a very good approximation for most experimentally relevant scenarios, including that we shall consider in more detail below. Alternatively, taking the limit $C \rightarrow \infty$ while $\gamma/\Omega$ is fixed, we find the form 
\begin{equation}
V_{X_+} = \frac{ \sqrt{ 4 \eta C (\bar{n}_{\rm th} + 1/2) + 1 } - 1 }{4\eta C} \rightarrow \frac{ \sqrt{ \bar{n}_{\rm th} + 1/2 } }{2 \sqrt{ \eta}} \frac{1}{\sqrt{C}} . \label{asymptotic2VXp}
\end{equation}
The scaling with $C$ is now the same as in the case $\Omega = 0$ (c.f. Eq.~(\ref{VXpOmegaZero})), which itself scales as in the case for single-mode back-action-evading measurement \cite{clerk:QND}. Clearly, for a sufficiently strong measurement (large cooperativity), the conditional variance of the measured collective quadrature observable, $\hat{X}_+$, is squeezed below the vacuum level ($1/2$) and tends asymptotically to zero. 

To assess whether we are truly generating a two-mode squeezed state and entanglement, we also need to check that the $\hat{P}_-$ quadrature is squeezed by the measurement (c.f. Eq.~(\ref{eq:Duan})). 
It can be shown, see App.~\ref{bestestimates}, that
\begin{eqnarray}
V_{P_-} & = & V_{X_+} + \left( \frac{\gamma}{\Omega} \right)^2 \frac{1}{16\eta C} \left[ 16\eta^2 C^2 \bar{n}_{\rm tot} \right. \nonumber \\
& & \left. + \left(1 + 4\eta C \bar{n}_{\rm tot} + \sqrt{1+8\eta C \bar{n}_{\rm tot} } \right)^2 \right] , \label{eq:VPmVXpPerturb}
\end{eqnarray}
to second-order in $\gamma/\Omega$, where we have introduced the notation for the total thermal \emph{and} quantum fluctuations,
\begin{equation}
\bar{n}_{\rm tot} = \bar{n}_{\rm th} + 1/2 . \label{eq:totalfluctuations}
\end{equation}
Note that Eq.~(\ref{eq:VPmVXpPerturb}) does not imply that the difference between $V_{P_-}$ and $V_{X_+}$ diverges as $C \rightarrow \infty$. In this limit higher-order contributions are important, and the perturbative result of Eq.~(\ref{eq:VPmVXpPerturb}) is not valid. However, provided the ratio $\gamma/\Omega$ is small, from Eq.~(\ref{eq:VPmVXpPerturb}) we do have 
\begin{equation}
V_{P_-} \sim V_{X_+}. \label{eq:VPmVXp}
\end{equation}
That is, in the limit that the collective oscillator frequency greatly exceeds the mechanical damping rate, the collective quadrature $\hat{P}_-$ that is dynamically coupled to the directly measured observable $\hat{X}_+$ is also effectively measured and therefore conditionally squeezed. Accordingly, the measurement conditionally generates a two-mode squeezed state. 

\setlength{\belowcaptionskip}{-5pt }

\begin{figure}[th]
\begin{center}
\includegraphics[scale=0.46]{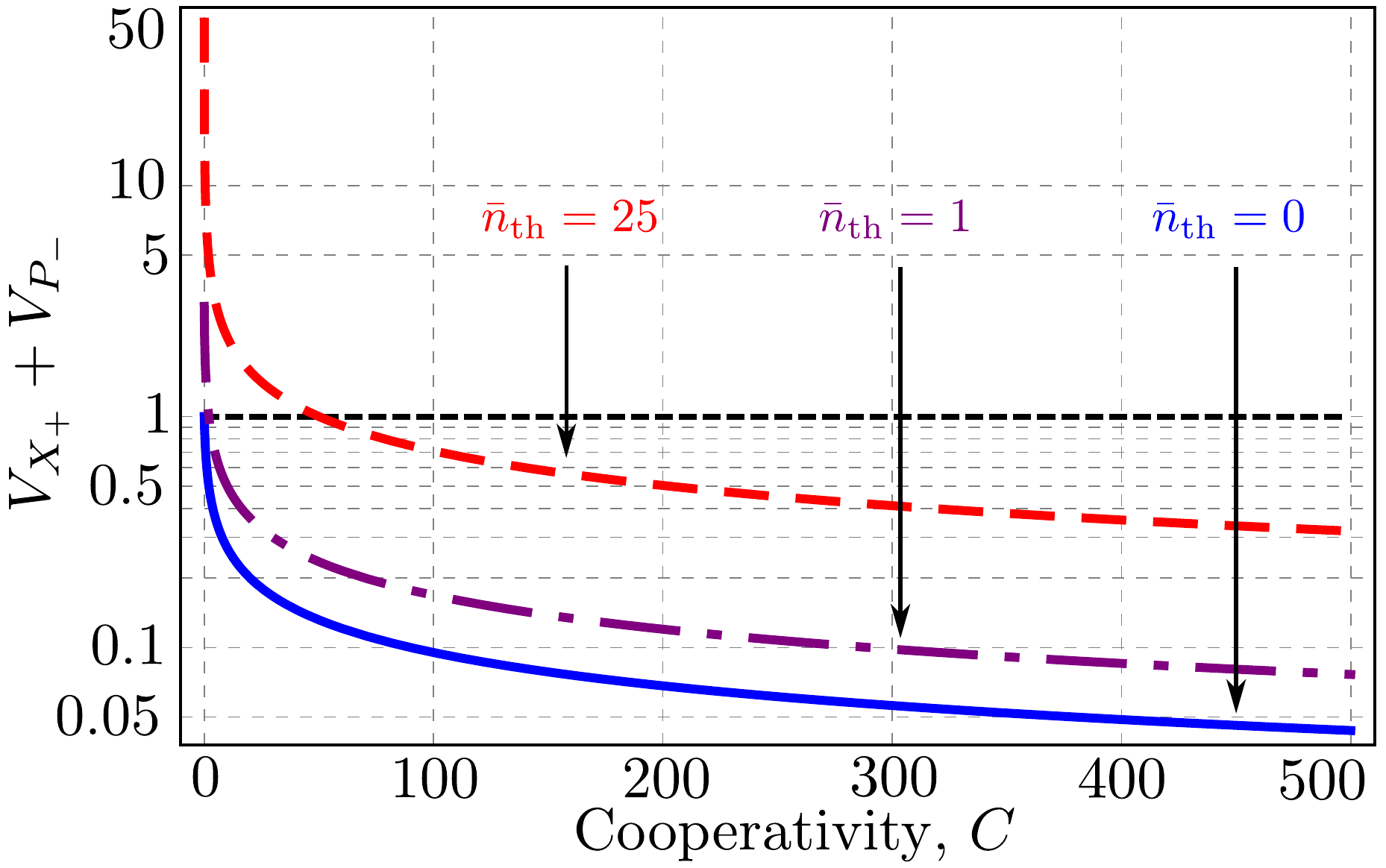}
\end{center}
\caption{ (Color online). The Duan quantity, $V_{X_+} + V_{P_-}$, as a function of the cooperativity $C$, for effective temperatures corresponding to $\bar{n}_{th}=0,1,25$, and in the regime $\Omega \gg \gamma$.  
The bound on the Duan criterion, as in Eq.~(\ref{eq:Duan}), is shown as the dashed black line. The quantity $V_{X_+} + V_{P_-}$ is seen to be below this line for sufficiently large cooperativities, indicating that the mechanical oscillators are entangled. This remains the case even for a relatively large thermal occupation. The required cooperativities are experimentally accessible. 
} 
\label{fig:V1}
\end{figure}

\setlength{\belowcaptionskip}{-10pt }

That both $\hat{X}_+$ and $\hat{P}_-$ are effectively measured (and therefore conditionally squeezed) in the regime $\Omega \gg \gamma$ can be seen in the following way. Defining primed operators by the transformation $\hat{Z}'_\pm = e^{-i\hat{H}_0 t} \hat{Z}_{\pm} e^{+i\hat{H}_0 t}$ with $\hat{H}_0 = \Omega ( \hat{X}_+ \hat{X}_- + \hat{P}_+ \hat{P}_- )$, we can express the observables $\hat{X}_+$ and $\hat{P}_-$ as
\begin{equation}
\left[ \begin{array}{c} \hat{X}_+ \\ \hat{P}_- \end{array} \right] = \left[ \begin{array}{cc} \cos \Omega t & \sin \Omega t \\ -\sin \Omega t & \cos \Omega t \end{array} \right] \left[ \begin{array}{c} \hat{X}'_+ \\ \hat{P}'_- \end{array} \right] . \label{eq:additionallyrotatingframe}
\end{equation}
While the observables $\hat{X}_+$ and $\hat{P}_-$ oscillate at $\pm \Omega$, the observables $\hat{X}'_+$ and $\hat{P}'_-$ are constants of the motion with respect to the closed system dynamics. From Eq.~(\ref{eq:additionallyrotatingframe}) it is clear that by continuously monitoring $\hat{X}_+$, we are continuously monitoring both $\hat{X}'_+$ and $\hat{P}'_-$. Since these observables also determine $\hat{P}_-$, we are effectively measuring $\hat{P}_-$ as well (provided that $\Omega \gg \gamma$).  

Recalling the ``subspaces'' (or subsystems) introduced in Sec.~\ref{sec:system}, we now know the conditional variances of the observables in the ``measured'' subsystem. For the observables in the ``perturbed'' subsystem, the variances in the limit $\gamma /\Omega \rightarrow 0$ follow from Eqs.~(\ref{eq:VXmndzero}) and (\ref{eq:VPpndzero}) as
\begin{equation}
V_{X_-}, V_{P_+} = \bar{n}_{\rm th} + 1/2 + C/2 . 
\end{equation}
Effectively, we are measuring both $\hat{X}_+$ and $\hat{P}_-$ in this limit, and so the conjugate observables of both are equally perturbed. 

\subsubsection{Entanglement}
Next we explicitly consider the entanglement of the mechanical oscillators. From Eqs.~(\ref{asymptoticVXp}) and (\ref{eq:VPmVXp}), in the regime $\Omega \gg \gamma$ and the strong measurement limit, we find that $V_{X_+} + V_{P_-} < 1$, implying that our mechanical oscillators are conditionally entangled. Substituting the asymptotic form of Eq.~(\ref{asymptoticVXp}) into Duan's criterion leads to the sufficient condition on the measurement strength for the generation of mechanical entanglement by measurement, 
\begin{equation}
C > \frac{2( \bar{n}_{\rm th} + 1/2 )}{\eta} . \label{eq:entanglementcriterion}
\end{equation}
Experimentally, this is not an overly demanding condition. It is interesting to note, however, that this condition does not exhibit the total insensitivity to temperature that has been found in a scheme to generate an EPR channel in a cascaded atomic and nanomechanical system \cite{hammerer:PRL}. The primary distinction is that here we are describing steady-state entanglement, rather than instantaneous entanglement arising from a strong single-shot feedback operation. 

\setlength{\abovecaptionskip}{-10pt }

\begin{figure}[th]
\begin{center}
\includegraphics[scale=0.4]{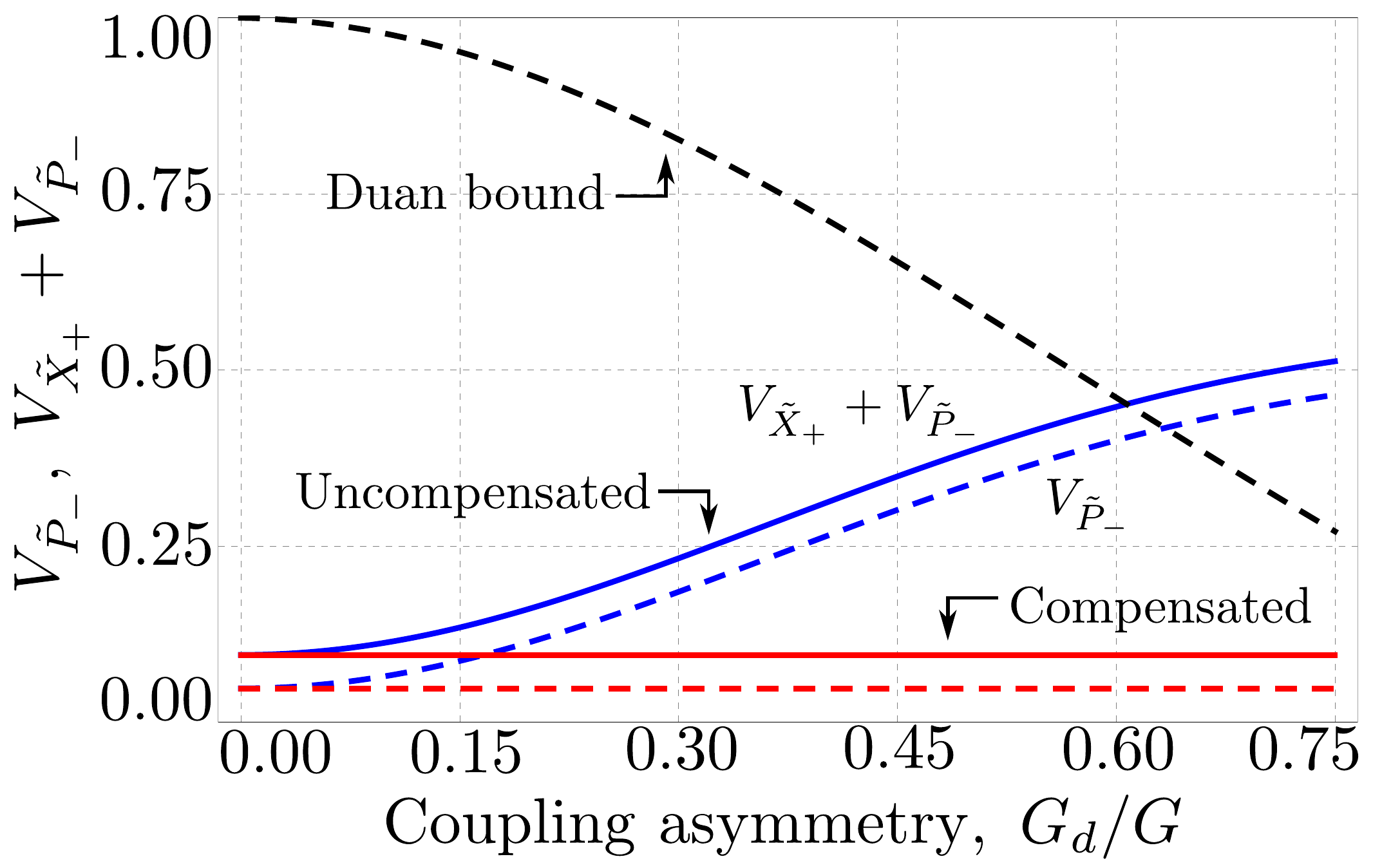}
\end{center}
\caption{ (Color online). The variances $V_{\tilde{P}_-}$ (dashed lines), and the generalized Duan quantities $V_{\tilde{X}_+} + V_{\tilde{P}_-}$ (solid lines) as a function of the optomechanical coupling asymmetry $G_d/G$, for the original (Eq.~(\ref{eq:Hamtilded})) and compensated (Eq.~(\ref{eq:HamtildedComp})) Hamiltonians. We assumed $\gamma_a = \gamma_b$, though even for large damping asymmetries, the modification to these lines is small. The bound on the generalized Duan inequality, as in Eq.~(\ref{eq:generalizedDuan}), is shown as the curved dashed black line. This plot was prepared with the effective occupation set to $\bar{n}_{\rm th} =0$ and a cooperativity of $C=100$. At such low temperature, the scheme is highly robust against coupling asymmetry. Even without compensation, the mechanical oscillators remain entangled for a coupling asymmetry of up to $\sim 60\%$. With compensation, the scheme is unaffected by coupling asymmetry. } 
\label{fig:V3}
\end{figure}

The Duan quantity of Eq.~(\ref{eq:Duan}) is plotted in Fig.~\ref{fig:V1}, as a function of cooperativity, for a range of thermal occupations and for parameters corresponding to the experiment of Teufel and co-workers \cite{teufel:groundstate}. It is seen that the mechanical oscillators will be entangled in an achievable parameter regime, even for mechanical oscillators initially in a thermal state far from the ground state. 

Further, one can explicitly evaluate the entanglement measure known as the logarithmic negativity \cite{vidal:lognegativity}. Using the result of Eq.~(\ref{asymptoticVXp}) we find
\begin{equation}
E_{\mathcal{N}} = \frac{1}{2} \left\{ { \rm log}_2 \left[ \eta C/ ( \bar{n}_{\rm th} + 1/2 ) \right] - 1 \right\} ,
\end{equation}
in the regime specified by the criterion of Eq.~(\ref{eq:entanglementcriterion}), and zero otherwise. The onset of entanglement, as a function of cooperativity, according to this measure is found to be consistent with that predicted by the application of Duan's criterion (c.f. Eq.~(\ref{eq:entanglementcriterion})).

\subsection{The Asymmetric Case}
If the damping rates or the optomechanical coupling rates of the mechanical oscillators are different, an analytical solution for the steady-state covariance matrix is no longer possible and we must resort to numerical methods. The one exception to this is the case where we have coupling asymmetry alone and compensation as per Eq.~(\ref{eq:HamtildedComp}). In this case, the asymptotic results of Eqs.~(\ref{asymptoticVXp}) and (\ref{asymptotic2VXp}) remain valid, implying entanglement can still be achieved for strong measurements; see App.~\ref{bestestimates} for more details. 

\begin{figure}[th]
\begin{center}
\includegraphics[scale=0.39]{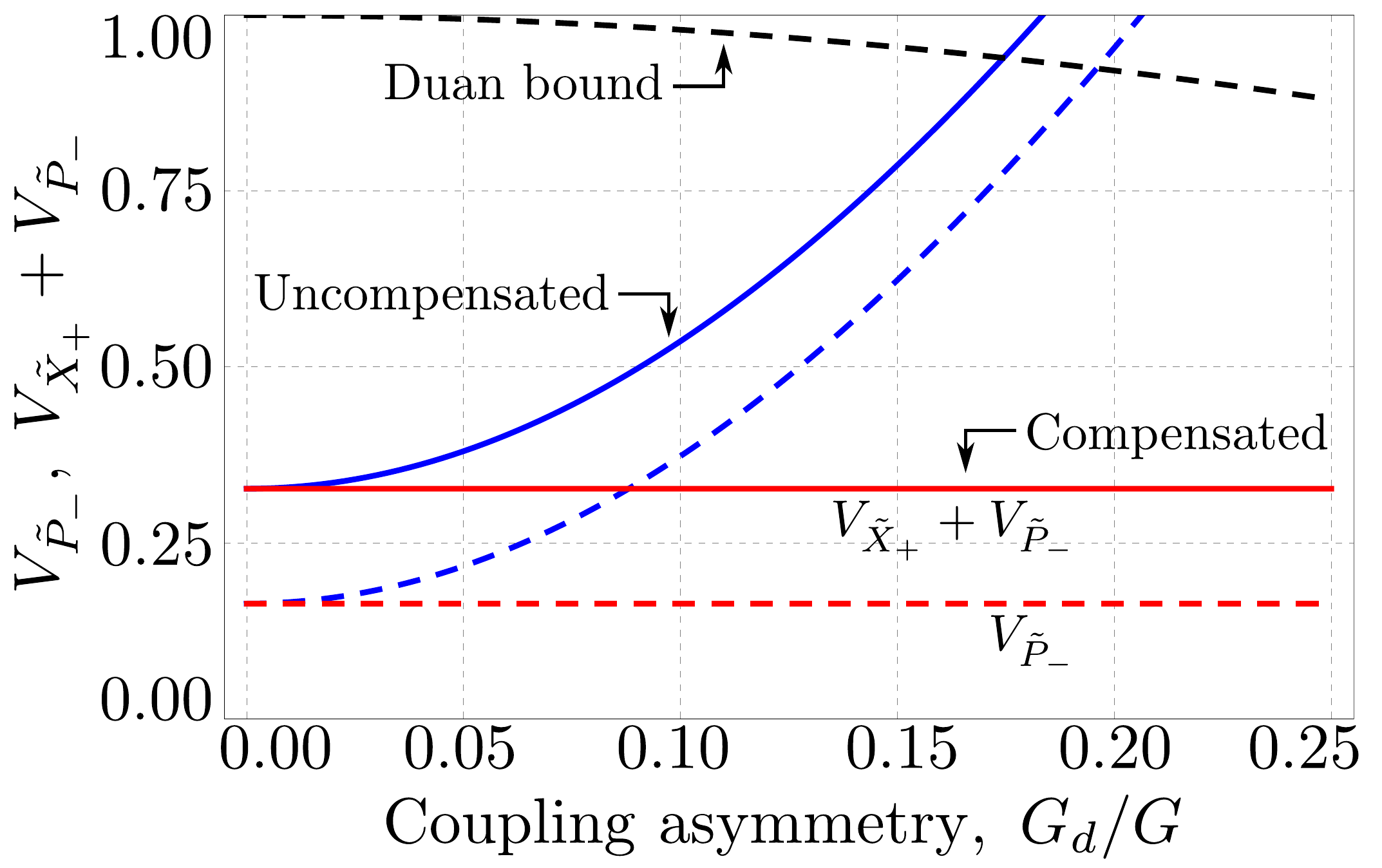}
\end{center}
\caption{ (Color online). The variances plotted in Fig.~\ref{fig:V3}, but now with an effective thermal occupation $\bar{n}_{\rm th} = 5$. At this higher effective temperature, the robustness of the scheme (without compensation) is reduced. The mechanical oscillators remain entangled for coupling asymmetries up to $\sim 17\%$. With compensation, the scheme is unaffected by coupling asymmetry.  } 
\label{fig:V2}
\end{figure}

Figs.~\ref{fig:V3} and \ref{fig:V2} show the steady-state conditional variance $V_{\tilde{P}_-}$ and the generalized Duan quantity $V_{\tilde{X}_+}+V_{\tilde{P}_-}$ as a function of the coupling asymmetry $G_d/G$. These are plotted assuming $\gamma_a = \gamma_b$, though the entanglement generated is only very weakly dependent on damping asymmetry, even when this asymmetry is large. Curves are shown for both the original and compensated cases in both figures, for the effective occupation $\bar{n}_{\rm th}=0$ in Fig.~\ref{fig:V3} and for $\bar{n}_{\rm th}=5$ in Fig.~\ref{fig:V2}. Note that in the original (uncompensated) case, the variance of the directly measured observable $V_{\tilde{X}_+}$ (the difference between the solid and dashed lines) is independent of the coupling asymmetry, while the variance of the dynamically coupled observable $V_{\tilde{P}_-}$ is less effectively squeezed as the coupling asymmetry is increased. 

At $\bar{n}_{\rm th} = 0$, the entanglement generation is seen to be highly robust against coupling asymmetry, even without compensation. The mechanical oscillators remain entangled up to a coupling asymmetry of $\sim 60\%$. At $\bar{n}_{\rm th} = 5$, the robustness against coupling asymmetry is reduced. In this case, the oscillators only remain entangled for a coupling asymmetry up to $\sim 17\%$. It should, however, be possible to independently cool the mechanical oscillators via auxiliary cavities such that they are initially at low thermal occupations \cite{teufel:groundstate}. Further, with compensation included, the entanglement generation is totally insensitive to coupling asymmetry. 
Furthermore, the coupling asymmetry for which entanglement generation is possible, beyond a threshold value, is not strongly dependent on the cooperativity.

\setlength{\abovecaptionskip}{-12.5pt }

\subsection{Experimental Parameters}
\label{parameterlist}

The experimental parameters used to prepare Figs.~\ref{fig:Spectra}-\ref{fig:V5} correspond to a micromechanical membrane in a superconducting microwave cavity, a design due to Teufel and co-workers \cite{teufel:groundstate}. One could conceivably form the microwave cavity from two micromechanical membranes, forming the three-mode optomechanical system that we have studied. However, the two-mode BAE scheme we have described could be performed in a number of cavity optomechanics implementations. 

For the purpose of numerical calculations, the cavity resonance frequency is taken to be $8\ {\rm GHz}$ and the mechanical resonance frequencies are centred around $10 \ {\rm MHz}$. The cavity damping rate is $\kappa/2\pi = 200 \ {\rm kHz}$, such that the system is operating well into the resolved-sideband regime. The mechanical damping rate is $\gamma/2\pi = 100 \ {\rm Hz}$. The zero-point fluctuations of the membrane are given by $\Delta x = 4 \ {\rm fm}$, such that the \emph{nominal} single-photon optomechanical coupling rate in Eq.~(\ref{eq:HamOne}) is $g_i/2\pi = 200 \ {\rm Hz}$. Further, suppose the cavity is driven such that its steady-state population at the driven sidebands is $10^6$ photons. This corresponds to a (nominal) effective optomechanical coupling rate in Eq.~(\ref{eq:HamUntilded}) that is $G/2\pi = 200 \ {\rm kHz}$. These parameters correspond to a cooperativity of $C=4\times 10^3$. This is a much larger cooperativity than is actually required for the scheme presented. In fact, such an optomechanical coupling rate would invalidate the adiabatic elimination performed in order to obtain Eq.~(\ref{eq:eliminatedME}). 

For the purpose of calculations here, we take a nominal effective coupling $G/2\pi = 70.7 \ {\rm kHz}$, corresponding to a cooperativity of $C=500$. Throughout this work, a quantum measurement efficiency of $\eta = 1$ has been assumed. Note that, from the perspective of entanglement generation, one can compensate for the reduced efficiency via a larger cooperativity (c.f. Eq.~(\ref{eq:entanglementcriterion})).  
These parameters are within the limits of validity of our analysis: recall that we assumed resolved-sideband operation ($\omega_a,\omega_b \gg \kappa$) in order to avoid spurious back-action heating of the measured observable, and the ``good measurement'' regime ($\kappa > \Omega,p\tilde{\Omega},\tilde{G}$) such that it was possible to perform the adiabatic elimination.

\subsection{Summary} 
The conditional dynamics of the system under homodyne detection of the field output from the cavity are described by Eq.~(\ref{eq:fullME}). In the adiabatic limit, in which the cavity responds rapidly to the mechanical motion, we have the simplified description of Eq.~(\ref{eq:eliminatedME}). This equation allows us to calculate the steady-state conditional variances of the collective mechanical quadratures (see Eq.~(\ref{eq:RiccatiEqn})), and then assess the entanglement of the mechanical oscillators using Duan's criterion (see Eq.~(\ref{eq:generalizedDuan})). It is found that the mechanical oscillators will be entangled for reasonable experimental parameters (see Fig.~\ref{fig:V1}), even in the presence of considerable asymmetries (see Figs.~\ref{fig:V3} and \ref{fig:V2}). 

\section{Unconditional Variances and Feedback}
\label{sec:unconditionalvariances}
The variances and covariances calculated in the previous section as solutions of Eq.~(\ref{eq:RiccatiEqn}) are \emph{conditional} quantities. However, the measured output noise spectrum of Eq.~(\ref{eq:measuredspectrum}) is an \emph{unconditional} quantity. This means it consists of both the fluctuations described by the best estimate of the conditional variance ($V_{\tilde{X}_+}$) \emph{and} the fluctuations in the best estimate of the observable itself ($\left\langle \bar{X}^2_+ \right\rangle$):
\begin{equation}
V^{\rm tot}_{\tilde{X}_+} = V_{\tilde{X}_+} + \left\langle \bar{X}^2_+ \right\rangle . \label{eq:UnconditionalAndConditional}
\end{equation}
We can define the total (unconditional) variances of the other collective quadratures in a similar fashion. From Eq.~(\ref{eq:UnconditionalAndConditional}), the fact that the conditional variance is squeezed below the vacuum level \emph{does not} necessarily imply that the unconditional variance will be also. Here we calculate the total variances, and describe how one can use feedback to reduce the unconditional variances to the conditional variances. 
 
\subsection{Unconditional Variances}

Fluctuations in the best estimates of the collective quadrature observables are described by Eq.~(\ref{eq:bestestimates}). The \emph{steady-state} fluctuations are then given by the solution of the Lyapunov equation, 
\begin{equation}
\mathbf{M} \bar{\mathbf{\Sigma}} + \bar{\mathbf{\Sigma}} \mathbf{M}^T = - \vec{Q} \vec{Q}^T, 
\end{equation}
where $\bar{\mathbf{\Sigma}}$ is a matrix of second moments of best estimates of collective quadrature observables, described in the ordered basis defined by Eq.~(\ref{eq:bestestimatesvec}), $(\bar{X}_+,\bar{P}_-,\bar{X}_-,\bar{P}_+)^T$. The matrix $\mathbf{M}$ is given by Eq.~(\ref{eq:MDefs}), (\ref{eq:rotatedsystem}) or (\ref{eq:compensatedsystem}), in the symmetric, uncompensated asymmetric or compensated asymmetric cases, respectively, while $\vec{Q}$ is given by Eq.~(\ref{eq:noiseinput}) in the symmetric case and by Eq.~(\ref{eq:originalN}) in both asymmetric cases.

In the fully symmetric case, and in the relevant regime $\Omega \gg \gamma$, 
we find the total unconditional variances to be
\begin{equation}
V^{\rm tot}_{P_-} \sim V^{\rm tot}_{X_+} = V_{X_+} \left( 1 + 2\eta C V_{X_+} \right) . \label{eq:unconditionalvariance}
\end{equation}
Substituting the first asymptotic form of Eq.~(\ref{asymptoticVXp}) into Eq.~(\ref{eq:unconditionalvariance}) we find that 
\begin{equation}
V^{\rm tot}_{P_-} = V^{\rm tot}_{X_+} = \bar{n}_{\rm th} + 1/2 . 
\end{equation}
This is precisely as expected; a back-action-evading measurement should neither heat nor cool the mechanical oscillators in the unconditional sense. The unconditional variances in the asymmetric case, both without and with compensation, are given in App.~\ref{sec:UnconditionalAsymmetric}. Note, however, that Eq.~(\ref{eq:unconditionalvariance}) remains valid (in terms of rotated observables) for the compensated case with no damping asymmetry. 

\subsection{Measuring Conditional Variances}
\label{sec:measurevariances}

The conditional variances may be obtained from measurements of the unconditional variances and the fluctuations in the best estimates, as per Eq.~(\ref{eq:UnconditionalAndConditional}). The unconditional variances may be obtained directly from the measured noise spectral density, or equivalently, from the measurement record. The unconditional variance of $\tilde{X}_+$ is obtained directly, while the unconditional variance of $\tilde{P}_-$ follows from looking at the quadratures of the measurement signal itself, as described around Eq.~(\ref{eq:additionallyrotatingframe}).  

The fluctuations in the best estimates of $\tilde{X}_+$ and $\tilde{P}_-$ may be obtained from the filter of Eq.~(\ref{eq:bestestimates}). Written out explicitly in terms of the measurement record of Eq.~(\ref{eq:measurementrecordincrement}), the filter is
\begin{equation}
\frac{d}{dt} \vec{ \bar{V} } = \mathbf{M} \cdot \vec{\bar{V}} + \sqrt{4\eta \Gamma} \left[ I(t) - \bar{X}_+ \right] \vec{Q} , \label{eq:filter}
\end{equation}
where $\mathbf{M}$, $\vec{Q}$ and $\vec{\bar{V}}$ are as specified following Eq.~(\ref{eq:bestestimates}), and $I(t)=dr/dt$ is called the measurement \emph{current}.
Of course, the filter of Eq.~(\ref{eq:filter}) is itself dependent on knowledge of the system's covariance matrix. We assume the system covariances have reached their steady-state values: that is, one should use the calculated values that follow from the steady-state solution of Eq.~(\ref{eq:RiccatiEqn}).  

In the fully symmetric case, the measured subsystem decouples from the perturbed subsystem. 
Then the filter of Eq.~(\ref{eq:filter}) can be recast, in the limit $\Sigma_{+-} \rightarrow 0$, as 
\begin{equation}
\left[ \begin{array}{c} \bar{X}_+ (t) \\ \bar{P}_- (t) \end{array} \right] = \sqrt{4\eta \Gamma} \ V_{X_+} e^{-\tilde{\gamma} t} \left[ \begin{array}{c} \cos \Omega t \\ - \sin \Omega t \end{array} \right] \ast I(t) , \label{eq:XpPmFilter}
\end{equation}
where the asterisk denotes convolution and we have introduced the notation $\tilde{\gamma} = \gamma + 4\eta \Gamma V_{X_+} $. The fluctuations in the best estimates may be obtained from the filtered measurement record of Eq.~(\ref{eq:XpPmFilter}). 

\subsection{Feedback}
\label{sec:useFB}

\subsubsection{Unconditional Variances}
As noted earlier, feedback may be employed to reduce the fluctuations in the best estimates of our observables, and so reduce the unconditional variances to conditional variances. In general this may be achieved by adding the damping terms $-(\alpha \gamma/2) \bar{X}_+$ and $-(\alpha \gamma/2) \bar{P}_-$ to the equations for $\bar{X}_+$ and $\bar{P}_-$ (respectively) to the filter of Eq.~(\ref{eq:filter}). This requires the application of (asymmetric) feedback forces to the two mechanical oscillators, given by 
\begin{eqnarray}
F_{a(b)}(t) & = & \frac{ \alpha \gamma }{ \sqrt{2} } \left( \cos \theta \mp \sin \theta \right) \bar{X}_+ \sin \omega_m t \nonumber \\
& & + \frac{\alpha \gamma}{\sqrt{2}} \left( \cos \theta \pm \sin \theta \right) \bar{P}_- \cos \omega_m t , \label{eq:FBAsymmetric}
\end{eqnarray}
in the laboratory frame. From the form of Eq.~(\ref{eq:filter}), we expect that $\bar{X}_+$ and $\bar{P}_-$ will themselves oscillate at $\pm \Omega$, and therefore the applied feedback force of Eq.~(\ref{eq:FBAsymmetric}) will in fact have significant weight at the two mechanical resonance frequencies. In most cases (except the case where there are significant coupling \emph{and} damping asymmetries) it suffices to apply the feedback component proportional to the estimate $\bar{X}_+$ alone. 

In order to calculate the variances under feedback we can use an equivalent classical description of the dynamics of our quantum system \cite{doherty:ctsmeasurement}, coupled to the filter giving the best estimates of the collective quadrature observables (and upon which the feedback will be based).  
The classical quantity representing the collective mechanical quadrature $\tilde{X}_+$ is denoted $\tilde{x}_+$ (and similarly for the other collective quadratures). The total unconditional variance of the measured observable, under feedback, is then $V^{\rm fb}_{\tilde{X}_+} = \left\langle \tilde{x}^2_+ \right\rangle $. 

Now we can write the measurement record increment in terms of such a classical representation as $dr = \tilde{x}_+ dt + dW/\sqrt{4 \eta \Gamma}$. We can also write the measurement record increment in terms of the best estimate of the measured observable as $dr = \bar{X}_+ dt + d\tilde{W}/\sqrt{4\eta \Gamma} $. 
Comparing these expressions leads to  
\begin{equation}
d\tilde{W} = \sqrt{4\eta \Gamma} \left( \tilde{x}_+ - \bar{X}_+ \right) dt + dW. \label{eq:WienerIncrements}
\end{equation}

Putting Eq.~(\ref{eq:WienerIncrements}) together with the system dynamics and filter from Eqs.~(\ref{eq:QLE}) and (\ref{eq:filter}), respectively, \emph{both} the conditional and unconditional dynamics can be described by the Ornstein-Uhlenbeck process,
\begin{equation}
\frac{d}{dt} \vec{Y} = -\mathbf{S} \cdot \vec{Y} + \mathbf{T} \cdot \frac{d \vec{W} }{dt} , \label{eq:FBeqns}
\end{equation}
where the system matrix $\mathbf{S}$ and the noise weighting matrix $\mathbf{T}$ are defined in Eqs.~(\ref{eq:Sblock}) and (\ref{eq:Tblock}) of App.~\ref{FeedbackAsymmetric}. The state vector $\vec{Y}$ and noise increment vector $d\vec{W}$ in Eq.~(\ref{eq:FBeqns}) are formed as 
\begin{equation}
\vec{Y} =\left[ \begin{array}{c} \vec{\bar{V}} \\ \hline \vec{v} \end{array} \right] , \ \ \ d\vec{W} = \left[ \begin{array}{c} d\vec{\bar{W}} \\ \hline d\vec{w} \end{array} \right] ,\label{eq:statenoisevec}
\end{equation}
where $\vec{\bar{V}}$ is the vector of best estimates of collective quadrature observables introduced in Eq.~(\ref{eq:bestestimatesvec}), $\vec{v}$ is the vector of classical representations of the collective quadratures,
\begin{equation}
\vec{v} = ( \tilde{x}_+, \tilde{p}_-, \tilde{x}_-, \tilde{p}_+ )^T ,
\end{equation}
$d\vec{\bar{W}}$ is a vector of Wiener increments corresponding to the measurement noise,
\begin{equation}
d\vec{\bar{W}} = dW \left( 1,1,1,1 \right)^T ,
\end{equation}
and $d\vec{w}$ is a vector of independent Wiener increments describing the noise associated with the coupling of the oscillators to their mechanical environments, 
\begin{equation}
d\vec{w} = ( dW_{\tilde{x}_+}, dW_{\tilde{p}_-}, dW_{\tilde{x}_-}, dW_{\tilde{p}_+} )^T.
\end{equation}

Now Eq.~(\ref{eq:FBeqns}) is, in general, an \emph{eight}-dimensional Ornstein-Uhlenbeck process, with the steady-state variances given by solutions of the Lyapunov equation,
\begin{equation}
\mathbf{S} \mathbf{\Xi} + \mathbf{\Xi} \mathbf{S}^T = \mathbf{T}\mathbf{T}^T , \label{eq:SSvariances}
\end{equation}
where $\mathbf{\Xi}$ denotes the steady-state covariance matrix in the ordered basis $\vec{Y}$. Solving Eq.~(\ref{eq:SSvariances}) provides all of the conditional and unconditional variances of interest. Note, however, that the matrices $\mathbf{S}$ and $\mathbf{T}$ themselves depend on the steady-state conditional variances, which must first be obtained by solving the algebraic Riccati equation giving the steady-state of Eq.~(\ref{eq:RiccatiEqn}).

\subsubsection{The Symmetric Case}
In the fully symmetric case, the equations for the measured subsystem ($\hat{X}_+,\hat{P}_-$) in Eq.~(\ref{eq:FBeqns}) decouple from those for the perturbed subsystem ($\hat{X}_-,\hat{P}_+$). Since $\theta =0$ now, the feedback force applied to both mechanical oscillators, from Eq.~(\ref{eq:FBAsymmetric}), should be
\begin{equation}
F(t)= \frac{\alpha \gamma}{\sqrt{2}} \bar{X}_+ \sin \omega_m t . \label{eq:FBSymmetric}
\end{equation}
The equations for the measured subsystem \emph{alone} can be written in the same form as Eq.~(\ref{eq:FBeqns}), with the appropriate matrices specified in App.~\ref{sec:FBMatricesSymmetric}. This now describes a \emph{four}-dimensional Ornstein-Uhlenbeck process, with steady-state solutions for the covariances again given by the solution of Eq.~(\ref{eq:SSvariances}). The reduced system size in this case facilitates a perturbative solution. Expanding in the (assumed) small parameter $(1+\alpha )^{-1}$, where $\alpha$ is the feedback gain in Eq.~(\ref{eq:FBSymmetric}), we find
\begin{subequations}
\begin{eqnarray} 
V^{\rm fb}_{X_+} & = & V_{X_+} + 4\eta C \frac{V^2_{X_+}}{ 1+\alpha } , \label{eq:varianceFB} \\
V^{\rm fb}_{P_-} & = & V_{P_-} + 4\eta C \frac{V^2_{X_+}}{ 1+\alpha }  \label{eq:variancePFB},
\end{eqnarray}
\end{subequations}
to first-order in $(1+\alpha )^{-1}$ and in the regime $\Omega \gg \gamma$. That is, in the limit of a large feedback gain $\alpha$, the unconditional variances reduce to the conditional variances. The conditional two-mode squeezing and entanglement calculated in Sec.~\ref{sec:conditionalvariances} can therefore be converted to \emph{unconditional} two-mode squeezing and entanglement via feedback. In the opposite limit of no feedback ($\alpha = 0$), we have $V^{\rm fb}_{X_+} = V^{\rm fb}_{P_-}  = \bar{n}_{\rm th} + 1/2$, as expected.  

Note that in the (opposite) case $\Omega = 0$, the result of Eq.~(\ref{eq:varianceFB}) is exact, consistent with the known result for single-mode back-action-evading measurement \cite{clerk:QND}. In this case we also have $V^{\rm fb}_{P_-} = V_{P_-}$, as the observable $\hat{P}_-$ is neither measured nor perturbed by the measurement.  

\setlength{\abovecaptionskip}{-10pt }

\begin{figure}[th]
\begin{center}
\includegraphics[scale=0.43]{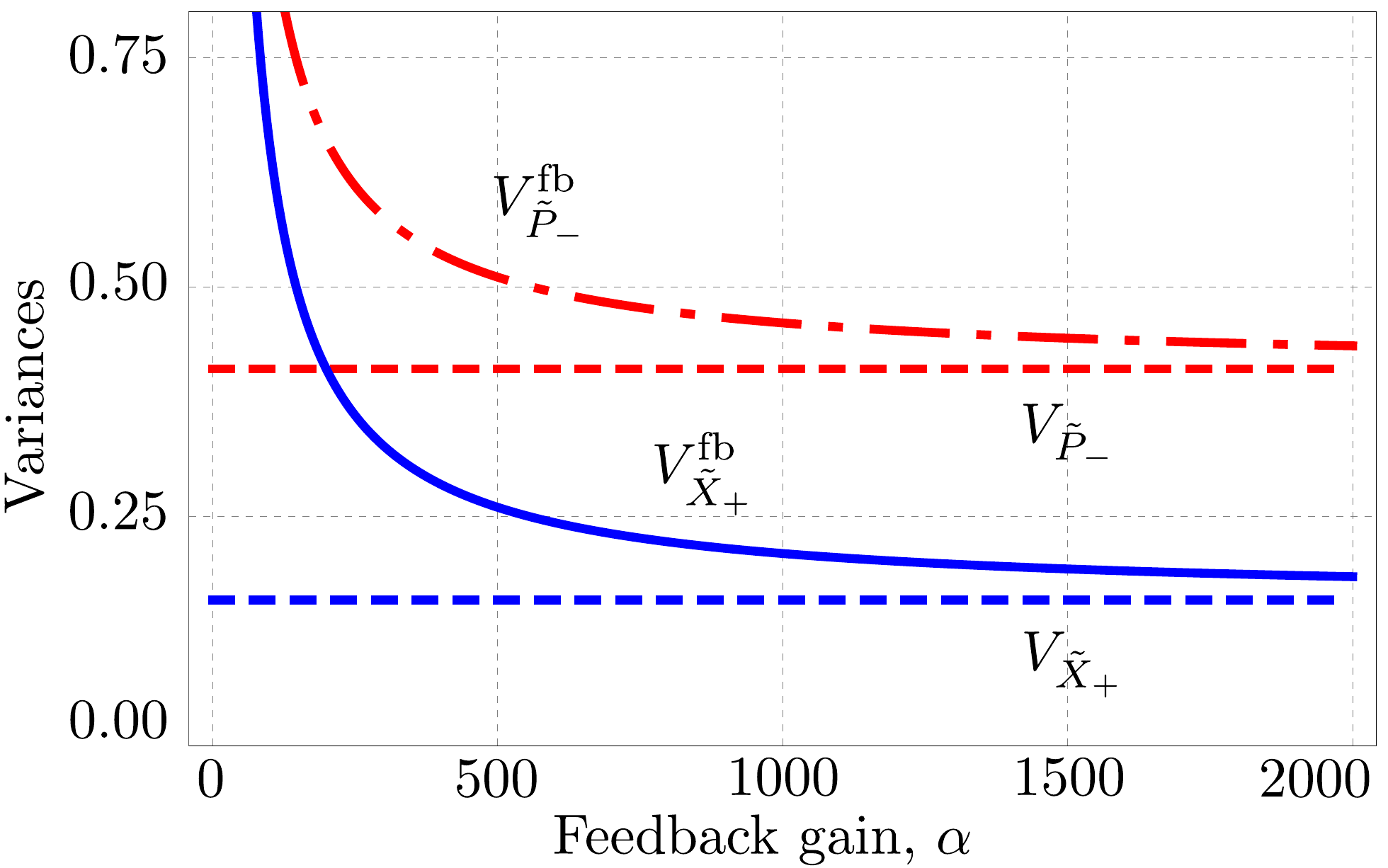}
\end{center}
\caption{ (Color online). The steady-state unconditional variances under feedback, $V^{\rm fb}_{ \tilde{X}_+ }$ and $V^{\rm fb}_{ \tilde{P}_- } $, as a function of the feedback gain $\alpha$ (introduced in Eq.~(\ref{eq:FBAsymmetric})), for an optomechanical coupling asymmetry of $G_d/G=0.05$, $\bar{n}_{\rm th} = 25$, $C=500$, and in the regime $\Omega \gg \gamma$. The steady-state conditional variances $V_{\tilde{X}_+}$ and $V_{\tilde{P}_-}$ are shown as the dashed horizontal lines. These conditional variances are unequal due to the coupling asymmetry. The steady-state unconditional variances approach the steady-state conditional variances asymptotically as the feedback gain is increased, though not as rapidly as in the case of symmetric coupling. } 
\label{fig:V5}
\end{figure}


\subsubsection{The Asymmetric Case}
Again, in the asymmetric case all conditional and unconditional covariances may be obtained by solving the system of Eq.~(\ref{eq:SSvariances}). However, the size of the system is such that we cannot obtain useful analytical solutions, and we resort to numerical solutions. Fig.~\ref{fig:V5} shows unconditional variances under feedback as a function of the feedback gain, in the uncompensated asymmetric case and for an optomechanical coupling asymmetry of $5\%$. It is seen that these steady-state unconditional variances ($V^{\rm fb}_{\tilde{X}_+}$ and $V^{\rm fb}_{\tilde{P}_-}$) approach the steady-state conditional variances ($V_{\tilde{X}_+}$ and $V_{\tilde{P}_-}$) asymptotically as the feedback gain is increased. However, this approach is slower than in the fully symmetric case. In the compensated asymmetric case, the unconditional variance $V^{\rm fb}_{\tilde{X}_+}$ is the same as $V^{\rm fb}_{\tilde{P}_-}$, and both tend to the (common) steady-state conditional variances under strong feedback.

\subsection{Summary}
In this section the unconditional variances of the collective mechanical quadratures were calculated. The fact that there is two-mode squeezing in the conditional variances does not imply that there is two-mode squeezing in the unconditional variances. However, feedback can be employed (see Eq.~(\ref{eq:FBAsymmetric})) to reduce the unconditional variances to their corresponding conditional variances (see Fig.~\ref{fig:V5}).
 

\section{Conclusions}
\label{sec:conclustions}
A back-action-evading measurement of a collective quadrature of two mechanical oscillators may be performed using a cavity detector. This is achieved, for a system operated in the resolved-sideband regime, by two-tone driving of the cavity with a detuning of plus and minus the average of the two mechanical oscillator frequencies. 

For the purpose of force sensing, one can surpass the full quantum limit on resonant force sensing provided the damping rate of the auxiliary oscillator is much lower than that of the driven oscillator, and the coupling asymmetry is matched accordingly. This would be experimentally challenging, so perhaps more useful is the fact that one can significantly surpass the standard quantum limit (and also surpass the full quantum limit) on detuned force sensing without such constraints on system asymmetries.

Further, in the adiabatic limit and in the regime where the collective mechanical oscillator frequency is much greater than the average mechanical damping rate, one can conditionally prepare an entangled two-mode squeezed state of the mechanical oscillators via measurement of the cavity output field. The presence of this entanglement may be verified in a straightforward manner from the measurement record. Further, simple feedback based on the measurement record may be used to convert the conditional two-mode squeezing and entanglement to unconditional entanglement.

Given the rapid progress in the measurement of microwave fields \cite{schwab,lang:CFs,lehnertrecent}, including demonstrations of feedback control \cite{siddiqi:FB,devoret:FB}, the proposed scheme appears to be a promising route towards the generation and verification of macroscopic, all-mechanical entanglement and force sensing beyond conventional quantum limits.

\section{Acknowledgements}
This work was supported by the DARPA ORCHID program, under a grant from the AFOSR. We thank Carlton Caves, Gerard Milburn, Warwick Bowen, Ian Petersen and Alexandre Blais for valuable discussions. 

\noindent
\emph{Note added:} After this paper was prepared, we became aware of a recent related work by Zhang \emph{et al.}, arXiv:1304.2459. This work described how the Hamiltonian of Eq.~(\ref{eq:HamUntilded}) in our paper could also be obtained for two condensates in an optical lattice. However, neither an analysis of the conditional dynamics under measurement nor a calculation of the force sensitivity of the system was given there. 



\appendix


\section{Derivation of Hamiltonian}
\label{HamiltonianDerivation}
We start from the Hamiltonian of Eq.~(\ref{eq:HamOne}), with the signal force given by Eq.~(\ref{eq:driving}). 
Moving into an interaction picture with respect to $\hat{H}_0 =  \omega_m ( \hat{a}^\dagger \hat{a} + \hat{b}^\dagger \hat{b} ) + \omega_c \hat{c}^\dagger \hat{c} $, we can rewrite this Hamiltonian as
\begin{eqnarray}
\mathcal{\hat{H}} & = & \Omega \left( \hat{a}^\dagger \hat{a} - \hat{b}^\dagger \hat{b} \right) + g_a \left( \hat{a} e^{-i \omega_mt} + \hat{a}^\dagger e^{i  \omega_m t} \right) \hat{c}^\dagger \hat{c} \nonumber \\ 
& & + g_b \left( \hat{b} e^{-i \omega_mt} + \hat{b}^\dagger e^{i  \omega_m t} \right) \hat{c}^\dagger \hat{c} + \hat{H}_{\rm diss} \nonumber \\
& & + \hat{H}_{\rm drive} . \label{eq:HamAppA}
\end{eqnarray}
The environments of the three oscillators are assumed to be ensembles of non-interacting oscillators. The usual Born, Markov and rotating-wave approximations are made on the system-environment interactions, and we also ignore environment-induced level shifts, as is typical for quantum optical master equations \cite{TMSS}. Therefore, the Heisenberg equations corresponding to Eq.~(\ref{eq:HamAppA}), neglecting noise terms, are
\begin{subequations}
\begin{eqnarray}
\dot{\hat{a}} & = & -i\Omega \hat{a} - ig_a e^{i \omega_m t} \hat{c}^\dagger \hat{c} - \frac{\gamma_a }{2} \hat{a} , \\
\dot{\hat{b}} & = & i\Omega \hat{b} - ig_b e^{i \omega_m t} \hat{c}^\dagger \hat{c} - \frac{\gamma_b}{2} \hat{b} , \\
\dot{\hat{c}} & = & -ig_a \left( \hat{a} e^{-i \omega_m t} + \hat{a}^\dagger e^{i \omega_m t } \right) \hat{c} \nonumber \\
& & - ig_b \left( \hat{b} e^{-i \omega_m t} + \hat{b}^\dagger e^{i \omega_m t} \right) \hat{c} \nonumber \\
& &  - i\mathcal{E}_+ e^{-i \omega_mt} - i\mathcal{E}_- e^{i \omega_m t} - \frac{\kappa}{2} \hat{c} .
\end{eqnarray}
\end{subequations}
Adopting the ansatz $ \hat{c}(t) = \hat{c}_0 (t) + \hat{c}_+(t) e^{-i\omega_m t} + \hat{c}_-(t) e^{i\omega_m t} $, assuming $\omega_m \gg \kappa$ (resolved-sideband regime), explicitly separating out the Fourier components of the field at the driven sidebands, and then equating frequency components \cite{woolley:squeezing}, we obtain the system:
\begin{subequations}
\begin{eqnarray}
\dot{\hat{a}} & = & -i\Omega \hat{a} - ig_a \left( \hat{c}^\dagger_0 \hat{c}_+ + \hat{c}^\dagger_- \hat{c}_0 \right) - \frac{\gamma_a}{2} \hat{a}, \label{eq:a} \\
\dot{\hat{b}} & = & i\Omega \hat{b} - ig_b \left( \hat{c}^\dagger_0 \hat{c}_+ + \hat{c}^\dagger_- \hat{c}_0 \right) - \frac{\gamma_b}{2} \hat{b}, \\
\dot{\hat{c}}_0 & = & -ig_a \left( \hat{a} \hat{c}_- + \hat{a}^\dagger \hat{c}_+ \right) - ig_b \left( \hat{b} \hat{c}_- + \hat{b}^\dagger \hat{c}_+ \right) \nonumber \\ 
& & - \frac{\kappa}{2} \hat{c}_0, \label{eq:c} \\
\dot{\hat{c}}_+ & = & -i\mathcal{E}_+ + i \omega_m \hat{c}_+  - \frac{\kappa}{2} \hat{c}_+ - ig_a \hat{a} \hat{c}_0 \nonumber \\ 
& & - ig_b \hat{b} \hat{c}_0 \label{eq:SBPlus}, \\
\dot{\hat{c}}_- & = & -i\mathcal{E}_- -i\omega_m \hat{c}_- - \frac{\kappa}{2} \hat{c}_- - ig_a \hat{a}^\dagger \hat{c}_0 \nonumber \\ 
& & -ig_b \hat{b}^\dagger \hat{c}_0 \label{eq:SBMinus}. 
\end{eqnarray}
\end{subequations}

Solving Eqs.~(\ref{eq:SBPlus}) and (\ref{eq:SBMinus}) for the steady-state at the driven sidebands, assuming the optomechanical couplings are relatively small, yields:
\begin{subequations}
\begin{eqnarray}
\left\langle \hat{c}_+ \right\rangle & = & \frac{ -i\mathcal{E}_+ }{ -i\omega_m + \kappa/2 } \equiv \bar{c}_+, \label{eq:zetaplus} \\
\left\langle \hat{c}_- \right\rangle & = & \frac{ -i\mathcal{E}_- }{ i \omega_m + \kappa/2 } \equiv \bar{c}_- . \label{eq:zetaminus}
\end{eqnarray}
\end{subequations}
Then we can write down the effective Hamiltonian, corresponding to Eqs.~(\ref{eq:a})-(\ref{eq:c}), as
\begin{eqnarray}
\mathcal{ \hat{H} } & = &  \Omega \left( \hat{a}^\dagger \hat{a} - \hat{b}^\dagger \hat{b} \right) + \hat{H}_{\rm diss} \nonumber \\
& & + g_a \left( \bar{c}_+ \hat{c}^\dagger \hat{a}^\dagger + \bar{c}^*_- \hat{c} \hat{a}^\dagger + \bar{c}^*_+ \hat{c} \hat{a} + \bar{c}_- \hat{c}^\dagger \hat{a} \right) \nonumber \\ 
& & + g_b \left( \bar{c}_+ \hat{c}^\dagger \hat{b}^\dagger + \bar{c}^*_- \hat{c} \hat{b}^\dagger + \bar{c}^*_+ \hat{c } \hat{b} + \bar{c}_- \hat{c}^\dagger \hat{b} \right) .
\end{eqnarray}
Now we assume that the cavity sideband amplitudes will have the same steady-state amplitudes but different phases, $\bar{c}_+ = \bar{c} e^{-i\psi}$ and $\bar{c}_- = \bar{c} e^{i\psi}$. Here $\bar{c}$ is assumed to be real, without loss of generality because it only changes the cavity quadrature to which the collective mechanical quadrature is coupled, and is readily compensated by adjusting a local oscillator phase. The Hamiltonian becomes:
\begin{eqnarray}
\mathcal{\hat{H}} & = & \Omega \left( \hat{a}^\dagger \hat{a} - \hat{b}^\dagger \hat{b} \right) + 2g_a \bar{c} \hat{X}_{a,\psi} \hat{X}_c + 2g_b \bar{c} \hat{X}_{b,\psi} \hat{X}_c \nonumber \\
& & + \hat{H}_{\rm diss} , \label{eq:intermediateHam}
\end{eqnarray}
where we have introduced quadratures of mechanical and electromagnetic modes as $\hat{X}_{q,\psi} \equiv (\hat{q} e^{i\psi} + \hat{q}^\dagger e^{-i\psi})/\sqrt{2}$ and $\hat{P}_{q,\psi} \equiv -i (\hat{q} e^{i\psi}-\hat{q}^\dagger e^{-i\psi})/\sqrt{2}$, with the lack of a phase subscript implying that the phase has been set to zero. The Hamiltonian of Eq.~(\ref{eq:intermediateHam}) may be rewritten as
\begin{eqnarray}
\mathcal{\hat{H}} & = & \Omega \left( \hat{X}_+ \hat{X}_- + \hat{P}_+ \hat{P}_- \right) + G\left( \cos \psi \hat{X}_+ - \sin \psi \hat{P}_+ \right) \hat{X}_c \nonumber \\
& & - G_d \left( \cos \psi \hat{X}_- - \sin \psi \hat{P}_- \right)  \hat{X}_c + \hat{H}_{\rm diss}  .
\end{eqnarray}
Irrespective of the value of $\psi$, there exists back-action in the ``unperturbed'' subspace due to asymmetric coupling ($G_d \neq 0$), and so we may set $\psi $ to zero. Then the Hamiltonian reduces to that given in Eq.~(\ref{eq:HamUntilded}).

\section{Heisenberg-Langevin Equations}
\label{sec:HLEquations}

The system of Heisenberg-Langevin equations describing the dynamics of the two mechanical oscillators is given in Sec.~\ref{sec:system} by Eq.~(\ref{eq:QLE}), repeated here for convenience,
\begin{equation}
\frac{d}{dt} \vec{V} = \mathbf{M} \cdot \vec{V} + \vec{F}_{\rm BA} + \mathbf{N} \cdot \vec{\xi}, \label{eq:QLErepeat}
\end{equation}
where $\vec{V}$ is the vector of collective mechanical quadrature operators of Eq.~(\ref{eq:vectoroperators}) and $\vec{\xi}$ is the vector of input noise operators of Eq.~(\ref{eq:noiseinput}). The coupling of the mechanical oscillators to the cavity mode is included in Eq.~(\ref{eq:QLErepeat}) as an inhomogeneity, the vector of back-action forces $\vec{F}_{\rm BA}$. In the fully symmetric case (symmetric mechanical damping rates and symmetric optomechanical coupling rates) the system matrix $\mathbf{M}$ is given by Eq.~(\ref{eq:MDefs}), the back-action force vector $\vec{F}_{\rm BA}$ is given by Eq.~(\ref{eq:FDefs}), and the input noise weighting matrix $\mathbf{N}$ is given by Eq.~(\ref{eq:noiseinput}). The Heisenberg-Langevin equations for the cavity quadratures themselves are given by Eqs.~(\ref{eq:Xcav}) and (\ref{eq:QLEPcav}). Here we give the Heisenberg-Langevin equations describing the system \emph{in the presence of both damping and coupling asymmetries}. These equations retain the form of Eq.~(\ref{eq:QLErepeat}) in all cases, and they shall be specified both in the original and rotated bases, and both without and with the compensating parametric driving. 


\subsection{Original System - Original Basis}


In the presence of both coupling and damping asymmetries, in the basis of the original collective observables of Eq.~(\ref{eq:vectoroperators}) and without compensation, the system matrix is
\begin{equation}
\mathbf{M} = \left[ \begin{array}{cccc} -\gamma/2 & \Omega & -d\gamma/2 & 0 \\ -\Omega & -\gamma/2 & 0 & -d\gamma/2 \\ -d\gamma/2 & 0 & -\gamma/2 & \Omega \\ 0 & -d\gamma/2 & -\Omega & -\gamma/2 \end{array} \right] , \label{eq:Moriginalbasis}
\end{equation}
where the average mechanical damping rate $\gamma$ and the dimensionless damping asymmetry  $d$ are as introduced in Eqs.~(\ref{eq:avedamping}) and (\ref{eq:dDefn}), respectively. 
Clearly from Eq.~(\ref{eq:Moriginalbasis}), the asymmetric damping directly couples the nominally ``measured'' $(\hat{X}_+,\hat{P}_- )$ and the nominally ``perturbed'' $(\hat{X}_-,\hat{P}_+)$ subsystems. The vector of noise input operators is given by Eq.~(\ref{eq:noiseinput}), though the noise input weighting matrix is
\begin{equation}
\mathbf{N} = \left[ \begin{array}{cccc} \sqrt{ \gamma_+ (d) } & 0 & \sqrt{ \gamma_- (d) } & 0 \\ 0 & \sqrt{ \gamma_+ (d) } & 0 & \sqrt{ \gamma_- (d) } \\ \sqrt{ \gamma_- (d) } & 0 & \sqrt{ \gamma_+ (d) } & 0 \\ 0 & \sqrt{ \gamma_- (d) } & 0 & \sqrt{ \gamma_+ (d) }  \end{array} \right] , \label{eq:originalN}
\end{equation}
where we have introduced the collective damping rate notation,
\begin{equation}
\sqrt{ \gamma_{\pm}(d) }  \equiv \sqrt{\gamma} \left( \sqrt{1+d} \pm \sqrt{1-d} \right)/2 .
\end{equation}
The vector of back-action forces associated with the coupling to the cavity is now 
\begin{equation}
\vec{F}_{\rm BA} = ( 0, G_d \hat{X}_c, 0, -G\hat{X}_c )^T, \label{eq:BAoriginalbasis}
\end{equation}
with the asymmetric coupling rate $G_d$ as introduced in Eq.~(\ref{eq:asymcoupling}). The equation for the cavity quadrature $\hat{X}_c$ is still Eq.~(\ref{eq:Xcav}), while the equation for $\hat{P}_c$ is now
\begin{equation}
\dot{\hat{P}}_c = - G \hat{X}_+ + G_d \hat{X}_- - \frac{\kappa}{2} \hat{P}_c + \sqrt{ \kappa } \hat{P}_{c,{\rm in}} . \label{QLE6un}   
\end{equation}
According to Eq.~(\ref{QLE6un}), the cavity is no longer coupled to (i.e. measures) $\hat{X}_+$ alone, but to a linear combination of $\hat{X}_+$ and $\hat{X}_-$. Accordingly, the observable $\hat{P}_-$ (which is dynamically coupled to the measured observable $\hat{X}_+$) is now heated by back-action from coupling to the cavity, as is also clear from Eq.~(\ref{eq:BAoriginalbasis}). Clearly, both types of asymmetries ruin the perfect back-action-evading property of the measurement scheme. The significance of these asymmetries is addressed throughout the paper. 


\subsection{Original System - Rotated Basis}
\label{originalsystemrotatedbasis}
In the presence of optomechanical coupling asymmetry it is convenient to move into a basis of \emph{rotated} collective quadrature observables, as defined in Eqs.~(\ref{eq:tilded0}) and (\ref{eq:tilded}). The Heisenberg-Langevin equations retain the form of Eq.~(\ref{eq:QLErepeat}), with $\vec{V}$ now denoting the vector of rotated quadrature observables as defined in Eq.~(\ref{eq:rotatedvector}), 
\begin{equation}
\vec{V} = ( \tilde{X}_+, \tilde{P}_-, \tilde{X}_-, \tilde{P}_+ )^T , 
\end{equation} 
and the vector of input noise operators as introduced in Eq.~(\ref{eq:tildenoise}),
\begin{equation}
\vec{\xi} = ( \tilde{X}_{+,{\rm in}} , \tilde{P}_{-,{\rm in}} , \tilde{X}_{-,{\rm in}} , \tilde{P}_{+,{\rm in}} )^T ,
\end{equation}
with the rotated collective input noise operators defined in terms of the original collective input noise operators in the obvious manner.

The system matrix is now, c.f. Eq.~(\ref{eq:Moriginalbasis}),
\begin{equation}
\mathbf{M} = \left[ \begin{array}{cccc} -\gamma/2 & \tilde{\Omega} & -d\gamma/2 & -p\tilde{\Omega} \\ -\tilde{\Omega} & -\gamma/2 & -p\tilde{\Omega} & -d\gamma/2 \\ -d\gamma/2 & p\tilde{\Omega} & -\gamma/2 & \tilde{\Omega} \\ p\tilde{\Omega} & -d\gamma/2 & -\tilde{\Omega} & -\gamma/2 \end{array} \right] , \label{eq:rotatedsystem}
\end{equation}
with $\tilde{\Omega}$ as introduced in Eq.~(\ref{eq:tildeOmegaDefn}), while the matrix $\mathbf{N}$ is still given by Eq.~(\ref{eq:originalN}).  
The back-action force vector is now
\begin{equation}
\vec{F}_{\rm BA} = ( 0 , 0 , 0 , -\tilde{G} \hat{X}_c )^T , \label{eq:tildeBA}
\end{equation}
while the Heisenberg-Langevin equations for the cavity quadratures are now Eqs.~(\ref{eq:Xcav}) and, c.f. Eq.~(\ref{QLE6un}),
\begin{eqnarray}
\dot{\hat{P}}_c  =  -\tilde{G} \tilde{X}_+ - \frac{\kappa}{2} \hat{P}_c + \sqrt{ \kappa } \hat{P}_{c,{\rm in}} . \label{QLE6}   
\end{eqnarray}
From Eq.~(\ref{eq:rotatedsystem}), it is clear that in the rotated basis both the damping and the coupling asymmetries directly couple the nominally measured $(\tilde{X}_+,\tilde{P}_- )$ and nominally perturbed $(\tilde{X}_-,\tilde{P}_+)$ subsystems. However, now only the rotated observable $\tilde{X}_+$ is coupled to the cavity (i.e. measured); indeed, this motivated the definition of the rotated observables in the first place. Accordingly, and as seen in Eq.~(\ref{eq:tildeBA}), the back-action of the cavity only directly heats $\tilde{P}_+$.

\subsection{Compensated System - Rotated Basis}
\label{compensatedsystemrotatedbasis}
Including compensating parametric driving as per Eq.~(\ref{eq:HamtildedComp}), the system matrix in the rotated basis of Eq.~(\ref{eq:rotatedvector}) is
\begin{equation}
\mathbf{M} = \left[ \begin{array}{cccc} -\gamma/2 & \tilde{\Omega} & -d\gamma/2 & 0 \\ -\tilde{\Omega} & -\gamma/2 & 0 & -d\gamma/2 \\ -d\gamma/2 & 2p\tilde{\Omega} & -\gamma/2 & \tilde{\Omega} \\ 2p\tilde{\Omega} & -d\gamma/2 & -\tilde{\Omega} & -\gamma/2 \end{array} \right] . \label{eq:compensatedsystem}
\end{equation}
The effect of the compensation is that the nominally measured subsystem is no longer coupled to the nominally perturbed subsystem via the coupling asymmetry, though it remains coupled through the damping asymmetry. This partial decoupling shall prove to be useful. The other parts of Eq.~(\ref{eq:QLErepeat}) remain unchanged: the vector of noise input operators is given by Eq.~(\ref{eq:tildenoise}), the vector of back-action forces is given by Eq.~(\ref{eq:tildeBA}), and the equations for the cavity quadratures are given by Eqs.~(\ref{eq:Xcav}) and (\ref{QLE6}).



\section{Calculation of Spectra}
\label{SpectraCalculation}

\subsection{Heisenberg-Langevin Equations in the Frequency Domain}
The back-action of the cavity on the mechanical oscillators may be determined by calculating noise spectra of the measured and perturbed observables. The system of Eq.~(\ref{eq:QLErepeat}) is readily solved in the frequency domain in all cases. Taking the Fourier transform of Eq.~(\ref{eq:QLErepeat}) leads to 
\begin{equation}
\vec{V} [\omega ] = - \mathbf{\chi} [\omega ] \cdot \mathbf{N} \cdot \vec{\xi }[\omega ] - \mathbf{\chi} [ \omega ] \cdot \vec{F}_{\rm BA} [\omega ] , \label{matrixsystem}
\end{equation}
where we have introduced the \emph{susceptibility matrix}, 
\begin{equation}
\mathbf{\chi} [ \omega ] = ( \mathbf{M} + i\omega \mathbf{1} )^{-1} . \label{eq:susceptibilitymatrix}
\end{equation}
For convenience, we also introduce the \emph{thermal} susceptibility matrix, 
\begin{equation}
\bar{\chi} [\omega ] \equiv \chi [\omega ] \cdot \mathbf{N}. \label{eq:thermalsusceptibility} 
\end{equation}
The symmetrized noise spectral densities of interest, for an observable $\hat{Z} = [ \vec{V} ]_i $, are defined in Eq.~(\ref{eq:symmnoise}).
They may be obtained from frequency-domain solutions of the Heisenberg-Langevin equations as
\begin{equation}
S_{Z} [ \omega ] = ( \hat{Z}[\omega ] )^\dagger \hat{Z} [\omega ] . \label{eq:spectrumfromsoln}
\end{equation}
From Eq.~(\ref{matrixsystem}), there are clearly two contributions to the noise spectra of the collective mechanical quadratures. The first term describes the intrinsic thermal and quantum fluctuations of the mechanical oscillators, while the second term describes the back-action heating of the mechanical oscillators due to their coupling to the cavity. We now calculate these contributions in turn. 

\subsection{Thermal Noise Spectra}
\label{thermalnoise}
The thermal noise spectra ultimately depend on the correlation functions of the input mechanical noise operators; these are
\begin{subequations}
\begin{eqnarray}
\langle \hat{Z}_{i,{\rm in}}(t) \hat{Z}_{i, {\rm in} } (t') \rangle & = & \frac{1}{2} \left( \bar{n}_a + \bar{n}_b + 1 \right) \delta (t-t') , \\
\langle \hat{Z}_{i, {\rm in}}(t) \hat{Z}_{j, {\rm in} } (t') \rangle & = & \frac{1}{2} \left( \bar{n}_a - \bar{n}_b \right) \delta (t-t') ,
\end{eqnarray}
\end{subequations}
for $i \neq j$ in the second correlation function, and where $\hat{Z} \in \{ \tilde{X}, \tilde{P} \}$ and $i,j \in \{ +,- \}$. Note that $\bar{n}_a ( \bar{n}_b )$ is the thermal occupation of the environment to which mechanical oscillator $a (b)$ is coupled. Therefore, the thermal contribution to the noise spectral density of the measured observable $\tilde{X}_+$ is
\begin{eqnarray}
S^{\rm th}_{\tilde{X}_+} [\omega ] & = & \frac{1}{2} \left( \bar{n}_a + \bar{n}_b + 1 \right) \left( \sum^4_{m=1} \left| \bar{\chi}_{1m} [\omega ] \right|^2 \right) \nonumber \\
& & \left. \left. + \frac{1}{2} \left( \bar{n}_a - \bar{n}_b \right) \left[ \sum^2_{m=1} \left( \bar{\chi}_{1m}^*[\omega ] \bar{\chi}_{1(m+2)}[\omega ] \right. \right. \right. \right. \nonumber \\ 
& & \left. \left. + \bar{\chi}_{1(m+2)}^*[\omega ]  \bar{\chi}_{1m} [\omega ] \right) \right] , \label{eq:thermalnoisespectraldensity}
\end{eqnarray} 
with $\bar{\chi}_{jk} [ \omega ]$ denoting elements of the thermal susceptibility matrix introduced in Eq.~(\ref{eq:thermalsusceptibility}). This quantity is evaluated explicitly, both at the resonant peaks ($\Omega$ in the original case, $\tilde{\Omega}$ in the compensated case) and at a large detuning $\Delta$, in Sec.~\ref{subsec:AsymmetricNoise}.

\subsection{Back-Action Noise Spectra}
\label{back-actionnoise}
The contribution to the noise spectral density of the measured observable due to back-action is itself dependent on the dynamics of the cavity. Accordingly, we first take the Fourier transform of the coupled cavity quadrature in Eq.~(\ref{eq:Xcav}),  
\begin{equation}
\hat{X}_c [ \omega ] = \chi_c [\omega ] \sqrt{\kappa} \hat{X}_{c, {\rm in} } [ \omega ] , \label{eq:XcavFrequency} 
\end{equation}
where we have introduced the \emph{cavity} susceptibility,  
\begin{equation}
\chi^{-1}_c [\omega ] \equiv -i\omega + \kappa/2. \label{eq:cavitysusceptibility} 
\end{equation}
The input noise correlation function for this cavity quadrature is
\begin{equation}
\langle \hat{X}_{c, {\rm in}} (t) \hat{X}_{c, {\rm in}} (t') \rangle = \frac{1}{2} \left( 2\bar{n}_c + 1 \right) \delta (t-t') , \label{eq:cavityinputnoise}
\end{equation}
where $\bar{n}_c$ is the thermal occupation of the cavity bath. Frequently, in experiments, $\bar{n}_c$ will be very close to zero. From Eqs.~(\ref{eq:tildeBA}), (\ref{matrixsystem}), (\ref{eq:XcavFrequency}), and (\ref{eq:cavityinputnoise}), the back-action contribution to the spectrum of the measured observable, assuming that $\kappa \gg \Omega$, is
\begin{equation}
S^{\rm ba}_{\tilde{X}_+} [\omega ] = \gamma \left| \chi_{14} [\omega ] \right|^2 \tilde{C} ( 2\bar{n}_c + 1 ) , \label{eq:BAbit}
\end{equation}
where $\chi_{14}[\omega ]$ is an element of the susceptibility matrix introduced in Eq.~(\ref{eq:susceptibilitymatrix}), and $\tilde{C}$ is the rotated cooperativity parameter introduced in Eq.~(\ref{eq:rotatedC}). Eq.~(\ref{eq:BAbit}) is evaluated explicitly \emph{at} resonance and \emph{far-detuned} from resonance in Sec.~\ref{subsec:AsymmetricNoise}. 

However, for the purpose of assessing the system's usefulness for force sensing, we need to know the bandwidth over which quantum limits can be surpassed. This requires knowledge of the full frequency-dependent contribution to the noise spectrum. In general this is complicated, though reasonably simple forms may be obtained in the perturbative regime (i.e. to second-order in $p$ and $d$), and in the extreme asymmetric case ($d=1$ and $p=-1$ without compensation, or just $d=1$ in the case with compensation).

\subsection{Output Noise Spectrum}
The total \emph{output} noise spectrum of the measured collective mechanical quadrature consists not only of the quantum and thermal fluctuations of the mechanical oscillators (calculated in App.~\ref{thermalnoise}) and the back-action heating due to the coupling to the cavity (calculated in App.~\ref{back-actionnoise}), but also includes the noise \emph{added by the detector} (i.e. the cavity). The total noise spectrum is readily calculated using the input-output formalism of quantum optics \cite{TMSS}, and this will facilitate comparisons with conventional quantum limits on measurement \cite{clerk:review}. 

To calculate this noise spectrum, we need to know how the collective mechanical quadrature couples to the cavity field. This is given, in the frequency domain, by the Fourier transform of Eq.~(\ref{QLE6}), 
\begin{equation}
\hat{P}_c[\omega ] = -\chi_c [\omega ] \tilde{G} \tilde{X}_+ [\omega ] + \chi_c[\omega ] \sqrt{\kappa} \hat{P}_{c, {\rm in}} [\omega ] , \label{eq:PcavFrequency}
\end{equation}
where the cavity susceptibility $\chi_c[\omega ]$ is defined in Eq.~(\ref{eq:cavitysusceptibility}). From Eqs.~(\ref{eq:XcavFrequency}) and (\ref{eq:PcavFrequency}), we have for the cavity mode annihilation operator,
\begin{equation}
\hat{c} [\omega ] = \sqrt{\kappa} \chi_c [\omega ] \hat{c}_{\rm in} [\omega ] - i\chi_c [ \omega ] \tilde{G} \tilde{X}_+[\omega ]/\sqrt{2} . 
\end{equation}
Applying the usual boundary condition for a single-sided optical cavity, 
\begin{equation}
\sqrt{\kappa} \hat{c} [ \omega ] = \hat{c}_{\rm in}[\omega ] + \hat{c}_{\rm out}[\omega ],
\end{equation}
we find the cavity output field to be
\begin{eqnarray}
\hat{c}_{\rm out} [\omega ] & = & - \bar{c}_{out} [\omega ] - \frac{ i (\omega - \omega_c ) + \kappa/2 }{ i(\omega - \omega_c ) - \kappa/2 } \hat{c}_{\rm in} [\omega ] \nonumber \\
& & - i \chi_c (\omega - \omega_c ) \sqrt{\kappa /2} \tilde{G} \tilde{X}_+ [ \omega - \omega_c ] ,
\end{eqnarray}
with the frequencies now specified in the (non-rotating) laboratory frame. Here $\bar{c}_{out} [ \omega ]$ describes the output field due to the coherent cavity driving fields, the second term describes fluctuations of the input field filtered by the cavity, and the third term carries the signal due to the mechanical oscillation. 

It is assumed that the output cavity field is subject to homodyne detection in the usual manner \cite{TMSS}. The measured homodyne current is then
\begin{eqnarray}
\hat{I} (t) & = & 
b^*_{\rm LO} (t) \hat{c}_{\rm out} (t) + b_{\rm LO} (t) \hat{c}^\dagger_{\rm out} (t) , \label{eq:homodynecurrent}
\end{eqnarray}
where the local oscillator amplitude is $b_{\rm LO}(t) = iB e^{-i\omega_c t}$, with $B$ assumed real without loss of generality. The Fourier transform of each component in Eq.~(\ref{eq:homodynecurrent}) is evaluated as a convolution integral, allowing us to calculate the spectrum of the homodyne measurement current $S_I [\omega ]$, as quoted in Eqs.~(\ref{eq:homodynenoise}) and (\ref{eq:measuredspectrum}) \cite{clerk:QND}. 
 
\section{Force Sensing Transfer Functions}
\label{sec:TFs}

The ability to perform a back-action-evading measurement of an oscillating mechanical observable suggests the possibility of force sensing beyond conventional quantum limits. Given that the noise spectral density of the measured observable has been calculated (see Sec.~\ref{back-action}), the next task is to calculate the \emph{transfer function} relating the signal force on one mechanical oscillator to the measured collective mechanical quadrature. This allows one to determine the noise added by the force sensing scheme, and facilitates a comparison with conventional quantum limits (see Sec.~\ref{sec:forcesensitivity}). These transfer functions are calculated here. 

Suppose that the mechanical oscillator $a$ is subject to the signal force $f(t)$, as per Eq.~(\ref{eq:DrivingLabFrame}). In a frame rotating at the \emph{average} mechanical frequency $\omega_m $ (i.e.~the same frame used to write the Hamiltonian in Eq.~(\ref{eq:HamUntilded})), the signal force is described by
\begin{equation}
\hat{H}_F = \bar{f} (t) \hat{a} + \bar{f}^* (t) \hat{a}^\dagger , \label{eq:DrivingRotatingFrame}
\end{equation}
where we have
\begin{equation}
\bar{f} (t) = f(t) e^{-i\omega_a t} e^{i\Omega t}. 
\end{equation}
The Hamiltonian driving of Eq.~(\ref{eq:DrivingRotatingFrame}) adds a vector of driving forces $\vec{F}(t)$ to the Heisenberg-Langevin equations of Eq.~(\ref{eq:QLE}), given by 
\begin{equation}
\vec{F}(t) = - \left[  \begin{array}{c} (\cos \theta - \sin \theta ) \ \mathrm{Im} \ \bar{f} (t) \\  (\cos \theta + \sin \theta ) \ \mathrm{Re} \ \bar{f} (t) \\  (\cos \theta + \sin \theta )\ \mathrm{Im} \ \bar{f} (t) \\  (\cos \theta - \sin \theta )\ \mathrm{Re} \ \bar{f} (t) \end{array} \right] , \label{eq:drivingforce}
\end{equation}
where $\theta$ is the angle describing the coupling asymmetry introduced in Eq.~(\ref{eq:xiDefn}). In order to evaluate the force sensing transfer function, we can neglect the noise terms and the coupling to the cavity in Eq.~(\ref{eq:QLE}), leaving the system of Heisenberg equations, 
\begin{equation}
\frac{d}{dt} \vec{V} = \mathbf{M} \cdot \vec{V} + \vec{F}(t) , \label{eq:EOM}
\end{equation}
where $\vec{V}$ is the vector of rotated quadrature observables in Eq.~(\ref{eq:rotatedvector}), and the system matrix $\mathbf{M}$ is given by Eq.~(\ref{eq:rotatedsystem}) or (\ref{eq:compensatedsystem}) for the uncompensated or compensated cases, respectively. The system of Eq.~(\ref{eq:EOM}) with the drive of Eq.~(\ref{eq:drivingforce}) is readily solved in the frequency domain, leading to the required transfer functions. These results are now given, for the cases of a signal force \emph{resonant} with the mechanical oscillator and for a signal force \emph{far-detuned} from the mechanical resonance frequency.

\subsection{Detection of mechanically resonant force}

First we consider a mechanically resonant signal force. For the uncompensated system, assume the signal force $f(t)$ is contained in a narrow bandwidth about the mechanical resonance frequency $\omega_a$. This implies that in the rotating frame $\bar{f}[\omega]$ is peaked at $\pm \Omega$.  For the compensated system we assume $f(t)$ is contained in a narrow bandwidth about $\omega'_a$, leading to $\bar{f}[\omega]$ being peaked at $\pm \tilde{\Omega}$. The transfer function between the signal force and the measured observable takes the form specified by Eqs.~(\ref{eq:ForceSusceptibility}) and (\ref{eq:sensingsusceptibility}). 

The transfer functions are modified in the presence of asymmetries by the function $g_r(p,d)$. For the uncompensated system, this modification is given by 
\begin{eqnarray}
g_{r} (p,d) & = & \frac{ \sqrt{1+p^2} (1+d+p)  -1 - d - dp - p^2  }{ 2 (1-d^2 + p^2 ) } \nonumber \\ 
& & \times \cosec \left( \frac{ \arctan p } {2} \right) . \label{eq:gar}
\end{eqnarray}
This factor is just $1$ for vanishing asymmetries ($p,d=0$), implying that $\hat{X}_+$ responds resonantly to the force as expected. 
In the absence of coupling asymmetry, we find $g_{r}(p=0,d)=1/(1+d)$. That is, the gain in Eq.~(\ref{eq:sensingsusceptibility}) is given by the \emph{damping rate of the driven oscillator}.
Accordingly, for $d\rightarrow -1$ (the damping rate of the driven oscillator goes to zero), the gain diverges. For the compensated system, 
the transfer function modification is given by
\begin{eqnarray}
g_{r}(p,d) & = & \frac{1}{1-d^2 +d^2p^2} \left[ (1-d) \cos \left( \frac{ \arctan p }{2} \right) \right. \nonumber \\
& & \left. - dp \sin \left( \frac{ \arctan p }{2} \right) \right] . \label{eq:gcr}
\end{eqnarray}

\subsection{Detection of mechanically non-resonant force}
The second case of interest is that in which the signal force is far-detuned (by an amount $\Delta \gg \gamma$) from the mechanical resonance frequency. The transfer function is now specified by Eqs.~(\ref{eq:ForceSusceptibility}) and (\ref{eq:modulationnonresonant}). The modifications to the transfer functions due to asymmetries are given by the function $g_n(p,d)$. In the uncompensated case this is 
\begin{equation}
g_{n} (p,d) 
= \frac{1+p-\sqrt{1+p^2}}{ 2\sqrt{1+p^2} } \cosec  \left( \frac{ \arctan p }{2} \right) , \label{eq:gao}
\end{equation}
while in the compensated case we find
\begin{equation}
g_{n} (p,d) = \cos \left(  \frac{\arctan p}{2} \right) . \label{eq:gco}
\end{equation}
In both cases the modulation is independent of the damping asymmetry, and so approaches $1$ for vanishing coupling asymmetry.
 
\section{Best Estimates of Quadratures and Conditional Variances - The Symmetric Case} 
\label{bestestimates} 

\subsection{Quadratures}
In the fully symmetric case (equal optomechanical coupling rates and mechanical damping rates), the measured subsystem ($\hat{X}_+,\hat{P}_-$) decouples from the perturbed subsystem ($\hat{X}_-,\hat{P}_+$) in Eqs.~(\ref{eq:filter}) and (\ref{eq:QLErepeat}). For the measured subsystem \emph{alone}, we can still write the filter, for instance, in the form of Eq.~(\ref{eq:bestestimates}), but now with the matrices
\begin{subequations}
\begin{eqnarray}
\mathbf{M} & = & \left[ \begin{array}{cc} -\gamma/2 & \Omega \\ -\Omega & -\gamma/2 \end{array} \right], \label{eq:MSymmetric} \\
\vec{Q} & = & \sqrt{4\eta \Gamma} \left[ \begin{array}{c} V_{X_+} \\ \Sigma_{+-} \end{array} \right] , \label{eq:QSymmetric}
\end{eqnarray}
\end{subequations} 
where $\Gamma$ is the measurement rate introduced in Eq.~(\ref{eq:measurementrate}). 

\subsection{Conditional Variances}
\label{VariancesSymmetric}
Also, in the fully symmetric case the system matrix $\mathbf{M}$ of Eq.~(\ref{eq:rotatedsystem}) simplifies to that given in Eq.~(\ref{eq:MDefs}), and the system of conditional covariance equations of Eq.~(\ref{eq:RiccatiEqn}) becomes, writing out the equations for the independent elements explicitly, 
\begin{subequations}
\begin{eqnarray}
\dot{V}_{X_+} & = & 2\Omega \Sigma_{+-} - \gamma V_{X_+} -  4\eta \Gamma V^2_{X_+} + \gamma \bar{n}_{\rm tot} , \label{eq:VXpSym} \\
\dot{V}_{P_-} & = & -2\Omega \Sigma_{+-} - \gamma V_{P_-} -  4\eta \Gamma  \Sigma^2_{+-} \nonumber \\
& & + \gamma \bar{n}_{\rm tot} , \label{eq:VPmSym} \\
\dot{V}_{X_-} & = & 2\Omega \Sigma_{-+} - \gamma V_{X_-} -  4\eta \Gamma \Sigma^2_{XX} + \gamma \bar{n}_{\rm tot} , \label{eq:VXmSym} \\
\dot{V}_{P_+} & = & -2\Omega \Sigma_{-+} - \gamma V_{P_+} -  4\eta \Gamma  \Sigma^2_{++} + \gamma \bar{n}_{\rm tot} \nonumber \\
& & + \Gamma , \label{eq:VPpSym} \\
\dot{\Sigma}_{+-} & = & -\Omega V_{X_+} + \Omega V_{P_-} - \gamma \Sigma_{+-} \nonumber \\
& & -  4\eta \Gamma V_{X_+} \Sigma_{+-}, \label{eq:CpmSym} \\
\dot{\Sigma}_{-+} & = & -\Omega V_{X_-} + \Omega V_{P_+} - \gamma \Sigma_{-+} \nonumber \\
& & -  4\eta \Gamma \Sigma_{XX} \Sigma_{++}, \label{eq:CmpSym} \\
\dot{\Sigma}_{++} & = & -\Omega \Sigma_{XX} + \Omega \Sigma_{PP} -\gamma \Sigma_{++} \nonumber \\
& & -  4\eta \Gamma \Sigma_{++} V_{X_+}, \label{eq:CpSym} \\
\dot{\Sigma}_{--} & = & -\Omega \Sigma_{XX} + \Omega \Sigma_{PP} - \gamma \Sigma_{--} \nonumber \\
& & -  4\eta \Gamma \Sigma_{+-} \Sigma_{XX}, \label{eq:CmSym} \\
\dot{\Sigma}_{XX} & = & \Omega \Sigma_{++} + \Omega \Sigma_{--} - \gamma \Sigma_{XX}  + \gamma \bar{n}_d \nonumber \\
& & -  4\eta \Gamma V_{X_+} \Sigma_{XX}, \label{eq:CXSym} \\
\dot{\Sigma}_{PP} & = & -\Omega \Sigma_{++} - \Omega \Sigma_{--} - \gamma \Sigma_{PP} + \gamma \bar{n}_d \nonumber \\ 
& & -  4\eta \Gamma \Sigma_{+-} \Sigma_{++} , \label{eq:CPSym} 
\end{eqnarray}
\end{subequations}
with $\bar{n}_{\rm tot}$, accounting for quantum and thermal fluctuations, as defined in Eq.~(\ref{eq:totalfluctuations}). From Eq.~(\ref{eq:VPpSym}), note that the observable that is perturbed by the measurement, $\hat{P}_+$, is heated at the measurement rate $\Gamma$, as required by Heisenberg's uncertainty principle. Further note that Eqs.~(\ref{eq:VXpSym}), (\ref{eq:VPmSym}) and (\ref{eq:CpmSym}), including the equations for the variances of most interest ($V_{X_+}$ and $V_{P_-}$), form a closed system and analytical steady-state solutions are readily found.  

First consider the simplest case, $\Omega = 0$. Writing the measurement rate in terms of the cooperativity parameter, $\Gamma = \gamma C$, we have
\begin{subequations}
\begin{eqnarray}
V_{X_+} & = & \frac{ \sqrt{ 4 \eta C (\bar{n}_{\rm th} + 1/2) + 1/4 } - 1/2 }{ 4\eta C } \nonumber \\
& \rightarrow & \frac{\sqrt{ \bar{n}_{\rm th} + 1/2 } }{ 2\sqrt{\eta} } \frac{1}{\sqrt{C}} , \label{VXpOmegaZero} \\
V_{P_+} & = & \bar{n}_{\rm th} + 1/2 + C, \label{VPmOmegaZero} \\
V_{X_-} & = & V_{P_-} = \bar{n}_{\rm th} + 1/2 .
\end{eqnarray}
\end{subequations}
The variance of the measured observable ($\hat{X}_+$) falls with a stronger measurement (higher cooperativity), while the conjugate observable ($\hat{P}_+$) is heated to a greater extent. The uncoupled observables ($\hat{X}_-,\hat{P}_-$) are unaffected by the measurement. Note that if we set the thermal occupation of each mechanical oscillator to be the same, the results of Eqs.~(\ref{VXpOmegaZero}) and (\ref{VPmOmegaZero}) are consistent with that for a back-action-evading measurement of a quadrature of a single mechanical oscillator \cite{clerk:QND}. 

However, for both generating entangled states and force sensing beyond quantum limits, we require that the collective mechanical oscillation frequency, $\Omega$, is non-zero. Solving for the steady-state of Eqs.~(\ref{eq:VXpSym}), (\ref{eq:VPmSym}) and (\ref{eq:CpmSym}) in this case leads to a fourth-order polynomial equation in $V_{X_+}$. This may be solved analytically, and expressions for $\Sigma_{+-}$  and $V_{P_-}$ then follow in turn. We find: 
\begin{widetext}
\begin{subequations}
\begin{eqnarray}
V_{X_+} & = & \frac{1 }{ 4\eta C } \left( -1 + \frac{1}{\sqrt{2} \gamma } \sqrt{ 8\eta  \gamma^2 C \bar{n}_{\rm tot} + \gamma^2 - 4\Omega^2 + \sqrt{ \gamma^2 + 4\Omega^2 } \sqrt{ \gamma^2 + 4\Omega^2 + 16 \eta \gamma^2 C \bar{n}_{\rm tot} }  } \right) , \label{eq:VXpSoln} \\
\Sigma_{+-} & = & \frac{\gamma}{2\Omega} \left( V_{X_+} + 4\eta C V^2_{X_+} - \bar{n}_{\rm tot} \right) , \label{eq:CpmExpl} \\
V_{P_-} & = & 2\bar{n}_{\rm tot} - V_{X_+} - 4\eta C V^2_{X_+}  + \eta C \left( \frac{\gamma}{\Omega} \right)^2 \left( \bar{n}_{\rm tot} - V_{X_+} - 4\eta C V^2_{X_+} \right)^2 ,\label{eq:VPmSoln}
\end{eqnarray}
\end{subequations}
\end{widetext}
with $\bar{n}_{\rm tot}$ given by Eq.~(\ref{eq:totalfluctuations}). The results of Eqs.~(\ref{eq:VXpSoln})-(\ref{eq:VPmSoln}) also apply to the compensated asymmetric system, with the understanding that the observables referred to are then the rotated observables, and we must replace parameters without tildes by the corresponding parameters with tildes. More useful results based on Eqs.~(\ref{eq:VXpSoln})-(\ref{eq:VPmSoln}) are given in Sec.~\ref{subsec:ConditionalVariances}. Note also that the second expression in Eq.~(\ref{VXpOmegaZero}) follows as a limit of Eq.~(\ref{eq:VXpSoln}) in the regime $\Omega \ll \gamma$. Further, substituting the full result for $V_{X_+}$ from Eq.~(\ref{eq:VXpSoln}) into the expression for $V_{P_-}$ in terms of $V_{X_+}$ from Eq.~(\ref{eq:VPmSoln}), and re-expressing the result in terms of $V_{X_+}$, we find the result of Eq.~(\ref{eq:VPmVXpPerturb}).

In the compensated case (with no damping asymmetry), the solutions for the steady-state covariances are given by Eqs.~(\ref{eq:VXpSoln})-(\ref{eq:VPmSoln}), with the understanding that they are now solutions for the rotated observables, and parameters without tildes must be replaced by the corresponding parameters with tildes. Consequently, the asymptotic results of Eqs.~(\ref{asymptoticVXp}) and (\ref{asymptotic2VXp}) remain valid.

Given the solutions in Eqs.~(\ref{eq:VXpSoln})-(\ref{eq:VPmSoln}), Eqs.~(\ref{eq:CpSym})-(\ref{eq:CPSym}) form a linear system 
\begin{equation}
\mathbf{A}_i \cdot \vec{R}_i = \vec{B}_{ij}, \label{eq:justlinearsystem}
\end{equation}
where the appropriate matrices are
\begin{subequations}
\begin{eqnarray}
\vec{R}_1 & = & (\Sigma_{++}, \Sigma_{--}, \Sigma_{XX}, \Sigma_{PP})^T , \label{eq:lin11} \\
\mathbf{A}_1 & = & \left[ \begin{array}{cccc} -\gamma & 0 & -\Omega & \Omega \\ 0 & -\gamma & -\Omega & \Omega \\ \Omega & \Omega & -\gamma & 0 \\ -\Omega & -\Omega & 0 & -\gamma \end{array} \right] \nonumber \\
& & - 4\eta \gamma C \left[ \begin{array}{cccc} V_{X_+} & 0 & 0 & 0 \\ 0 & 0 & \Sigma_{+-} & 0 \\ 0 & 0 & V_{X_+} & 0 \\ \Sigma_{+-} & 0 & 0 & 0 \end{array} \right] , \label{eq:Aone} \\
\vec{B}_{11} & = & \left[ \begin{array}{c} 0 \\ 0 \\ -\gamma \bar{n}_d \\ -\gamma \bar{n}_d \end{array} \right], \label{eq:symmetriclinear1} \\ 
\vec{B}_{12} & = & \vec{B}_{11} - 2p\tilde{\Omega} \left[ \begin{array}{c} V_{X_+} \\ V_{P_-} \\ \Sigma_{+-} \\ \Sigma_{+-} \end{array} \right] , \label{eq:compensatedlinear1}
\end{eqnarray}
\end{subequations}
where $\vec{B}_{11}$ or $\vec{B}_{12}$ is the appropriate inhomogeneity in the symmetric or compensated case, respectively.
Subsequently, Eqs.~(\ref{eq:VXmSym}), (\ref{eq:VPpSym}) and (\ref{eq:CmpSym}) form a linear system of the form of Eq.~(\ref{eq:justlinearsystem}), now with
\begin{subequations}
\begin{eqnarray}
\vec{R}_2 & = & ( V_{X_-}, V_{P_+}, \Sigma_{-+} )^T , \label{eq:lin21} \\
\mathbf{A}_2 & = & \left[ \begin{array}{ccc} -\gamma & 0 & 2\Omega \\ 0 & -\gamma & -2\Omega \\ -\Omega & \Omega & -\gamma \end{array} \right] , \label{eq:Atwo} \\
\vec{B}_{21} & = & \left[ \begin{array}{c} 4\eta \gamma C \Sigma^2_{XX} - \gamma \bar{n}_{\rm tot} \\ 4\eta \gamma C \Sigma^2_{++} -\gamma \bar{n}_{\rm tot} - \gamma C \\ 4\eta \gamma C \Sigma_{XX} \Sigma_{++}  \end{array} \right] , \label{eq:Btwo} \\
\vec{B}_{22} & = & \vec{B}_{21} - 2p\tilde{\Omega} \left[ \begin{array}{c} 2\Sigma_{--} \\ 2\Sigma_{++} \\ \Sigma_{XX} + \Sigma_{PP} \end{array} \right] , \label{eq:lin23}
\end{eqnarray}
\end{subequations}
again with $\vec{B}_{21}$ or $\vec{B}_{22}$ being the appropriate inhomogeneity in the symmetric or compensated case, respectively.
Note that in the compensated case, as compared with the fully symmetric case, there are simply additional inhomogeneous terms, leading to excess heating in the perturbed subsystem.  

The solutions to these linear systems take simple forms in the fully symmetric case, particularly in the physically relevant case $\bar{n}_d = 0$ (both oscillators at the same temperature). The linear systems defined by Eqs.~(\ref{eq:lin11})-(\ref{eq:compensatedlinear1}) and (\ref{eq:lin21})-(\ref{eq:lin23}) are readily solved in turn, leading to
\begin{equation}
\Sigma_{++}, \Sigma_{--}, \Sigma_{XX} , \Sigma_{PP} = 0 , \label{eq:covariancesndzero}
\end{equation}
and subsequently to
\begin{subequations}
\begin{eqnarray}
V_{X_-} = \bar{n}_{\rm th} + \frac{1}{2} + C \frac{2 \Omega^2}{ \gamma^2 + 4\Omega^2 },  \label{eq:VXmndzero} \\ 
V_{P_+} = \bar{n}_{\rm th} + \frac{1}{2} + C \frac{\gamma^2 + 2\Omega^2}{\gamma^2 + 4\Omega^2} , \label{eq:VPpndzero} \\ 
\Sigma_{-+} = C \frac{ \gamma }{\Omega} \frac{\Omega^2}{\gamma^2 + 4\Omega^2} . \label{eq:Cmpndzero}
\end{eqnarray}
\end{subequations}
In the limit $\gamma \gg \Omega$, $V_{X_-} \sim \bar{n}_{\rm th} + 1/2$ and $V_{P_+} \sim \bar{n}_{\rm th} + 1/2 + C$. In this limit, the measurement is of $\hat{X}_+$ alone, and so the conjugate observable $\hat{P}_+$ is heated by an amount corresponding to the cooperativity. In the opposite (and more interesting) limit, $\Omega \gg \gamma$, we find $V_{X_+} \sim V_{P_-} \sim \bar{n}_{\rm th} + 1/2 + C/2$. In this case, $\hat{X}_+$ is directly monitored, but it is dynamically coupled to $\hat{P}_-$. Therefore, both of these collective quadratures are effectively measured, and the corresponding conjugate quadratures are heated by an equal amount.  

\section{Unconditional Variances and Feedback}

\label{FeedbackAsymmetric}

\subsection{Unconditional Variances}
\label{sec:UnconditionalAsymmetric}
The expressions for the unconditional variances, in the absence of feedback, still take reasonably simple forms in the asymmetric case, provided we consider the regime where $\Omega \gg \gamma$. In the original (uncompensated) case we find, 
assuming that the asymmetries are small, 
\begin{subequations}
\begin{eqnarray}
\langle \tilde{X}^2_+ \rangle & = & 2\eta C V^2_{\tilde{X}_+}  \frac{ (1+p^2)(1+dp) - d^2/2 }{ (1+p^2) (1-d^2 + p^2 ) } , \label{eq:XtildeFluctSimple} \\
\langle \tilde{P}^2_- \rangle & = & 2\eta C V^2_{\tilde{X}_+} \frac{ 1+p^2 -d^2/2 }{ (1+p^2) (1-d^2 + p^2 )  }  . \label{eq:PtildeFluctSimple}
\end{eqnarray}
\end{subequations}
Note that there is a discrepancy between Eqs.~(\ref{eq:XtildeFluctSimple}) and (\ref{eq:PtildeFluctSimple}) only when there is \emph{both} a coupling asymmetry and a damping asymmetry.

\subsection{Feedback - The General Case}

The equation describing both the best estimates of the collective mechanical quadratures \emph{and} the classical representations of these observables takes the form of Eq.~(\ref{eq:FBeqns}). In the general (asymmetric) case, the measured and perturbed subspaces are coupled, such that the state vector contains all collective quadratures and the noise vector includes measurement noise and the noise due to the independent mechanical environments, as specified by Eq.~(\ref{eq:statenoisevec}).
The system matrix in Eq.~(\ref{eq:FBeqns}) may be expressed in the block matrix form
\begin{equation}
\mathbf{S} \equiv \left[ \begin{array}{c|c} \mathbf{M} + \mathbf{F} + \mathbf{Q}_1 & -\mathbf{Q}_1 \\ \hline \mathbf{F} & \mathbf{M} \end{array} \right] \label{eq:Sblock}
\end{equation}
where (as usual) $\mathbf{M}$ is given by Eq.~(\ref{eq:MDefs}), (\ref{eq:rotatedsystem}) or (\ref{eq:compensatedsystem}), $\mathbf{F}$ is the feedback matrix defined in block matrix form as
\begin{equation}
\mathbf{F} \equiv \left[ \begin{array}{c|c} (\alpha \gamma /2) \mathbf{I}_2 & \mathbf{0}_{22} \\ \hline \mathbf{0}_{22} & \mathbf{0}_{22} \end{array} \right] ,
\end{equation}
with $\mathbf{I}_n$ denoting the $n\times n$ identity matrix, $\mathbf{0}_{mn}$ representing an $m \times n$ zero matrix, and 
\begin{equation}
\mathbf{Q}_1 \equiv \sqrt{4\eta \Gamma} \left[ \begin{array}{c|c} \vec{Q} & \mathbf{0}_{43} \end{array} \right] ,
\end{equation}
with $\vec{Q}$ as defined in Eq.~(\ref{eq:measurementnoiseweight}). Further, the input noise weighting matrix in Eq.~(\ref{eq:FBeqns}) may be expressed in block matrix form as
\begin{equation}
\mathbf{T} \equiv \left[ \begin{array}{c|c} \mathbf{Q}_2 & \mathbf{0}_{44} \\ \hline \mathbf{0}_{44} & \sqrt{ \gamma \bar{n}_{\rm tot} } \mathbf{I}_4 \end{array} \right] , \label{eq:Tblock}
\end{equation}
where we have introduced
\begin{equation}
\mathbf{Q}_2 \equiv \sqrt{4\eta \Gamma} \ \rm{Diag} \left[ V_{X_+}, \Sigma_{+-}, \Sigma_{XX}, \Sigma_{++} \right] .
\end{equation}

\subsection{Feedback - The Symmetric Case}
\label{sec:FBMatricesSymmetric}
In the fully symmetric case we need only consider the observables in the measured subsystem, such that Eq.~(\ref{eq:FBeqns}) describes a four-dimensional Ornstein-Uhlenbeck process. The state vector is then
\begin{equation}
\vec{Y} = \left( \bar{X}_+, \bar{P}_-, x_+, p_- \right)^T ,
\end{equation}
the input noise vector is 
\begin{equation}
d\vec{W} = \left( dW, dW, dW_{x_+} , dW_{p_-} \right)^T . \label{eq:Wiener}
\end{equation}
The system matrix is now, explicitly, 
\begin{equation}
\mathbf{S} \equiv \left[ \begin{array}{cccc} \gamma/2 + \alpha \gamma /2 + 4\eta \Gamma V_{X_+} & -\Omega & -4\eta \Gamma V_{X_+} & 0 \\ \Omega + 4\eta \Gamma \Sigma_{+-} &\gamma/2 & -4\eta \Gamma \Sigma_{+-} & 0 \\ \alpha \gamma/2 & 0 & \gamma/2 & -\Omega \\ 0 & 0 & \Omega & \gamma/2 \end{array} \right] , \label{eq:Ssmall}
\end{equation}
and the input noise matrix is now, explicitly,  
\begin{equation}
\mathbf{T} \equiv \rm{Diag} \left[ \sqrt{ 4\eta \Gamma } V_{X_+}  , \sqrt{4\eta \Gamma} \Sigma_{+-},  \sqrt{ \gamma \bar{n}_{\rm tot} } , \sqrt{ \gamma \bar{n}_{\rm tot} } \right] . \label{eq:Tsmall}
\end{equation}
Of course, both Eqs.~(\ref{eq:Ssmall}) and (\ref{eq:Tsmall}) could be expressed in block matrix form, as in Eqs.~(\ref{eq:Sblock}) and (\ref{eq:Tblock}).

\end{document}